\documentclass[reprint,aps,prl]{revtex4-2}

\usepackage{amsmath,amssymb,amsfonts,dsfont}
\usepackage{graphicx}
\usepackage{color}
\usepackage{hyperref}
\usepackage{subfigure}
\usepackage{dcolumn}
\usepackage{bm}
\usepackage{float}
\usepackage{mathtools}
\usepackage{amsthm}
\usepackage{bbm}
\usepackage{pxfonts}
\usepackage{pstricks}
\usepackage{verbatim}

\definecolor{orange}{cmyk}{0,0.5,1,0}
\definecolor{rossoCP3}{cmyk}{0,.88,.77,.40}
\definecolor{graa}{rgb}{0.8,0.8,0.8}
\definecolor{blaa}{rgb}{0.2,0.2,0.6}
\hypersetup{
	colorlinks, 
	bookmarksopen, 
	bookmarksnumbered,
	citecolor=blaa, 		
	linkcolor=rossoCP3,	
	urlcolor=rossoCP3,			
	    }

\newcommand{\beq}{\begin{eqnarray}}
\newcommand{\eeq}{\end{eqnarray}}

\newcommand{\SU}{\mathrm{SU}}
\newcommand{\Sp}{\mathrm{Sp}}
\newcommand{\SO}{\mathrm{SO}}
\newcommand{\U}{\mathrm{U}}

\begin{document}
\title{Selection toolkit for standard and non-standard GUTs}

\author{Giacomo Cacciapaglia}
\email{cacciapa@lpthe.jussieu.fr}
\affiliation{Laboratoire de Physique Theorique et Hautes Energies {\color{rossoCP3}LPTHE}, UMR 7589, Sorbonne Universit\'e \& CNRS, 4 place Jussieu, 75252 Paris Cedex 05, France.}
\author{Konstantinos Kollias}
\email{kkollias@clipper.ens.psl.eu}
\affiliation{Laboratoire de Physique Theorique et Hautes Energies {\color{rossoCP3}LPTHE}, UMR 7589, Sorbonne Universit\'e \& CNRS, 4 place Jussieu, 75252 Paris Cedex 05, France.}
\author{Aldo Deandrea}
\email{deandrea@ip2i.in2p3.fr} 
\affiliation{Universit{\'e} Claude Bernard Lyon 1, {\color{rossoCP3}IP2I} UMR 5822, CNRS/IN2P3, 4 rue Enrico Fermi, 69622 Villeurbanne Cedex, France;\\
Department of Physics, University of Johannesburg,
PO Box 524, Auckland Park 2006, South Africa.}
\author{Francesco Sannino}
\email{sannino@qtc.sdu.dk}
\affiliation{{\color{rossoCP3}{$\hbar$}QTC} \& the Danish Institute for Advanced Study {\color{rossoCP3}\rm{Danish IAS}},  University of Southern Denmark, Campusvej 55, DK-5230 Odense M, Denmark;}
\affiliation{Dept. of Physics E. Pancini, Universit\`a di Napoli Federico II, via Cintia, 80126 Napoli, Italy;}
\affiliation{INFN sezione di Napoli, via Cintia, 80126 Napoli, Italy}
\affiliation{Scuola Superiore Meridionale, Largo S. Marcellino, 10, 80138 Napoli, Italy,}

\begin{abstract}
Under a reasonable set of ab-initio assumptions, we define and chart the atlas of simple gauge theories with families of fermions whose masses are forbidden by gauge invariance. We propose a compass to navigate the atlas based on counting degrees of freedom. When searching for Grand-unification Theories with three matter generations, the free energy singles out the SU(5) Georgi-Glashow model as the minimal one, closely followed by SO(10) with spinorial matter. The atlas also defines the dryland of grand-unifiable gauge extensions of the standard model. We further provide examples relevant for gauge dual completions of the standard model as well as extensions by an additional SU(N) gauge symmetry.
\end{abstract}

\maketitle

\section{Introduction}

The Standard Model (SM) of particle physics \cite{Glashow:1961tr,Weinberg:1967tq,Salam:1968rm} is a remarkable construction, both in its theoretical content and in its success to provide an accurate description of the fundamental forces, excluding gravity. It is largely based on the gauge principle, introducing interactions via local symmetries \cite{Yang:1954ek}: $\SU(3)$ for the strong chromomagnetic ones and $\SU(2) \times \U(1)$ for the electroweak ones. This structure has been experimentally tested up to energies of few TeV and below the percent level in most sectors, with the energy frontier currently met at the Large Hadron Collider (LHC). Nevertheless, the SM must be considered as a low-energy effective theory, due to its theoretical shortcomings and the observed phenomena beyond its realms (such as the baryon asymmetry in the universe, neutrino masses and inflation).  However, when considering higher-energy models, it is not at all granted that their low-energy limit uniquely yields the SM. Usually, a selection is made in terms of matter content, gauge groups and symmetry-breaking patterns following the primer of low-energy physics. A tantalizing example is given by Grand-Unification Theories (GUTs), where the semi-simple gauge symmetry of the SM is replaced by a more minimal structure, with the final goal of describing interactions via a single gauge coupling at higher energies.
The first attempt to unify quarks and leptons was made by means of a fourth color \cite{Pati:1974yy}, whereas it soon became clear that the complete SM structure could be derived from $\SU(5)$ \cite{Georgi:1974sy} or $\SO(10)$ \cite{Fritzsch:1974nn}. The exceptional group $E_6$ \cite{Gursey:1975ki} offers a popular alternative, mainly motivated by heterotic string theory \cite{Ito:2010df}.

Inspired by GUTs, in this article we propose to start, instead, from a general atlas of gauge theories characterized by a single gauge coupling. These \emph{grand-unified theories} differ from GUTs as they are not required to contain the SM  (see \cite{Zee:1980ce} for an early attempt to the latter). We propose using the atlas to chart the dryland of grand-unifiable models that extend the SM gauge symmetry. We put forward three concrete application. Firstly, we map the GUT landscape to the atlas by identifying the minimal theoretical requirements that may lead to a selection of the SM at low energies. Secondly, we apply the atlas to SM duals, i.e. theories that do not contain the SM but rather flow to it at low energies. Finally, we consider theories that contain additional gauge factors to the SM ones at low energies.

\vspace{0.3cm}

Remaining within the realm of quantum field theory in four-dimensional space-time, one can reasonably expect that, below the Planck scale, any theory of Nature features a number of Weyl fermions and, at least, one gauge group. This is in line with the observed particle spectrum of the SM, which is constituted predominantly by fermions and gauge fields. Furthermore, we follow the principle that any mass that is not forbidden by a local symmetry exists and is as large as possible.
Hence, we will only consider fermions whose mass is forbidden by gauge invariance, because they are naturally light compared to any higher scale such as the Planck one. Note that we do not consider (light) scalar fields, as their masses cannot be forbidden. This reasoning can be extended to a supersymmetric (SUSY) framework, with Weyl fermions and gauge bosons replaced by chiral and vector superfields in $\mathcal{N}=1$ \cite{Salam:1974yz,Ferrara:1974ac} or to non-supersymmetric gauge-Yukawa theories featuring ultraviolet fixed points of either completely safe \cite{Litim:2014uca,Litim:2015iea,Sannino:2015sel,Pelaggi:2017wzr,Abel:2017ujy,Fabbrichesi:2020svm,Bednyakov:2023asy} or free nature \cite{Cheng:1973nv,Callaway:1988ya,Holdom:2014hla,Giudice:2014tma,Pica:2016krb}.

Henceforth, we define the \emph{Grand-unified Theory Atlas} (GTA) as being populated by gauge theories that meet the following requirements:
\begin{itemize}
\item[1)] One non-abelian simple Lie group as gauge symmetry. In this article we consider $\SU(N)$, $\SO(N)$, $\Sp(2N)$ and exceptional gauge groups.
\item[2)] Anomaly-free fermionic matter families. By a \emph{family} we mean an irreducible set of massless fermions so that a mass term is forbidden by gauge invariance \footnote{For vector-like theories, a complete classification was first presented in \cite{Dietrich:2006cm}.}.
\item[3)] Asymptotic freedom in the gauge coupling, which is the only renormalizable interaction, in the absence of scalars.
\end{itemize}
The requirement 1) stems from the simplest realization of a single gauge coupling. Alternatively, one could consider products of identical groups related by cyclical symmetries, in the spirit of trinification \cite{Glashow:1984gc,Babu:1985gi}. An atlas of these models will be presented in a forthcoming publication.
Asymptotic freedom in 3), instead, is crucial in order to obtain well-defined theories in the ultra-violet, which allows us to characterize their properties. We'll see a concrete use of this property in the first application of the atlas below.

We will restrict our considerations to $d=4$ space-time and renormalizable theories, where asymptotic freedom allows the theory to be in principle viable at arbitrarily high energies. 
Each non-equivalent theory is characterized by a number $N_g$ of families (a.k.a. generations), either identical or not, limited by asymptotic freedom.
In our GTA, irreducible families are defined in two complementary ways:
\begin{itemize}
    \item \emph{Chiral families} are made of gauge anomaly-free combinations of complex representations \cite{Eichten:1985ft}. They emerge from several representations of $\SU(N)$ with $N \geq 3$ \footnote{Complex anomaly-free irreducible representations exist for $n\geq 5$ \cite{Eichten:1982pn,Gripaios:2024lmr}, however they always violate asymptotic freedom.}, the spinorial of $\SO(10+4k)$ with $k\in \mathbb{N}$, and the fundamental of the exceptional group $E6$. Here, multiple identical families are allowed.
    \item \emph{Pseudo-real families}, made of a single representation whose (Majorana) mass is forbidden.  They correspond to the spinorial of $\SO(11+8k)$, $\SO(12+8k)$ and $\SO(13+8k)$ with $k\in \mathbb{N}$, some representations of $\SU(N)$ and $\Sp(2N)$, and the fundamental of $E_7$. For pseudo-real representations of $\Sp(2N)$ groups, one needs to ensure the absence of the $\pi_4$ Witten anomalies \cite{Witten:1982fp,Wang:2018qoy}, hence combinations of two representations may be necessary.
\end{itemize} 
Pseudo-real families only appear as a single copy, since, for an even number, mass terms can be written by pairing them.
The low-energy properties of theories with pseudo-real families remain uncertain \cite{Witten:1982fp} and they will be studied in a forthcoming publication~\footnote{G.~Cacciapaglia, K.~Kollias, F.~Sannino, in preparation.}. A complete classification of $\SU(N)$ and $\Sp(2N)$ families can be found in the appendix. We remind the reader that we do not impose on the theories in the GTA to lead to the SM at low energies, contrary to GUTs. Detailed phenomenological considerations, therefore, are beyond the scope of the GTA definition, and they should be considered as part of a further step towards a proper use of the atlas.

Scalar fields may play an important role in phenomenology, as they may be required in order to break the gauge group and generate fermion masses via Yukawa couplings. Details depend on the usage of the GTA. Scalars are less constrained by the atlas criteria, as they do not contribute to anomalies and their mass is unconstrained. Furthermore, the presence of additional Yukawa and quartic couplings change the high-energy behavior of the theory. The inclusion of scalars should be considered as adding a third dimension to the atlas, which should be studied case-by-case depending on how the atlas is used.

In the following we will present three complementary applications of the GTA.

\section{Finding the Standard Model}

Starting from the GTA at very high energies, one can hope to find a selection rule that gives a strong preference for a GUT model that may lead to the SM at low energies.
Outside the atlas, one could also envision gravity-free, asymptotically safe GUTs, but these typically require a more involved high energy dynamics \cite{Bajc:2016efj,Cacciapaglia:2020qky}. GUTs within phenomenological string models have also been considered \cite{Chen:2010tg,Ito:2010df,Kobayashi:2004ya}, where a unified dynamics emerges at the string scale.

To our purpose, we need a criterion to rank the theories in the GTA. It may be tempting to count the degrees of freedom in each family \cite{Zee:1980ce}, defined via the number of Weyl spinors
\begin{equation}
    n_f = \sum_{r \in f} d_r\,,
\end{equation}
where $d_r$ is the dimension of the representation $r$, and the sum spans over the representations that define a family. In Table~\ref{tab:ng1}, we list the first few families, ranked by increasing $n_f$. However, to minimize the choice of the gauge group and matter representation simultaneously, the question of the relative weight between bosons and fermions becomes essential. A simple ansatz would consist in the naive counting of degrees-of-freedom without relative weight. Instead, we prefer to rely on meaningful physical quantities, which are well-defined in quantum field theory. One such example is based on using the free energy of the system as a way to count the degrees-of-freedom of the theory \cite{Appelquist:1999vs,Appelquist:2000qg}. As the theories are asymptotically free at high energy, the associated free energy can be computed as the one of a relativistic gas of free particles at high temperature. In this regime, the free energy $ \mathcal{F}$, normalized to the temperature, reads: 
\begin{equation}
        f_{FE} = - \frac{45 \mathcal{F}}{\pi^2 T^4} = d_G + \frac{7}{8} \sum_f n_g^f n_f\, ,
    \end{equation}
where $d_G$ is the dimension of the adjoint representation, and $n_g^f$ counts the copies of the irreducible family $f$ (so that $N_g \equiv \sum_f n_g^f$). For SUSY theories, the analogous quantity reads
\begin{equation}
    f_{FE}^S = \frac{15}{8} \left( d_G + \sum_f n_g^f n_f \right)\,.
\end{equation}
The values of the free energy counters are listed in the left-hand columns of Table~\ref{tab:ng1} for models of one family. 

One can also envision using another scalar quantity to minimize with respect to gauge and matter content. To compare with free energy, we will also consider the degree-of-freedom count $f_a$ arising from the $a-$function \cite{Cardy:1988cwa}. In the asymptotically free regime of the theories, it can be perturbatively computed and reads \cite{Dondi:2017civ}
  \begin{equation}
        f_a = (4\pi)^2 
        {a} =  \frac{124 d_G + 11 \sum_f n_g^f n_f}{720}\,.
    \end{equation}
The above expression indicates that the function $f_a$ increases the contribution of the gauge bosons compared to the free-energy counterpart. Note that both $f_a$ and $f_{FE}$ have been used in the literature to study the flow of theories from the ultra-violet to the infra-red \cite{Appelquist:1999hr,Appelquist:1999vs,Antipin:2013pya}.    

In Table~\ref{tab:ng1}, we recognize renown examples of chiral theories: the Georgi-Glashow ones based on fundamental and two-index antisymmetric starting with $\SU(5)$ \cite{Georgi:1974sy,Appelquist:1999vs} and the Bars-Yankielowicz ones based on fundamental and two-index symmetric starting with $\SU(3)$ \cite{Bars:1981se,Appelquist:1999vs}. Both counters favor theories with the smallest possible gauge group, hence preferring the two $\Sp(4)$ and $\SU(3)$ theories over the $\SU(5)$ one. The same ranking occurs in the SUSY cases, with some theories already excluded by the loss of asymptotic freedom (indicated by a zero in the max $n_g^f$ column). However, note that the same ranking may change for models with more than one generation.
\begin{table}[h]
	\centering
	\begin{tabular}{l|c|cccc|cc}
	Group & Family & Type & $n_f$  & $\xi_f$ & $n_g^f \leq$ & $f_{FE}$  & $f_a$ \\\hline
    $\SU(5)$ & ${\bf 10}_A + {\bf \overline{5}}_F$ & C & $15$ & $4/55$ & $13\ (7)$ & $37.1\ (73.1)$ & $4.36$\\
    $\Sp(4)$ & ${\bf 16}$ & PR & $16$ & $12/33$ & $1\ (1)$ & $24\ (48.7)$ & $1.97$ \\
    $\SO(10)$ & ${\bf 16}_{\rm Spin}$ & C & $16$ & $4/88$ & $21\ (11)$ & $59\ (114.)$ & $7.99$\\  
    $\SU(6)$ & ${\bf 20}_{A3}$ & PR & $20$ & $6/66$ & $1\ (1)$ & $52.5\ (103.)$ & $6.33$ \\
    $\Sp(6)$ & ${\bf 6}_F + {\bf 14}_{A3}$ & PR & $20$ & $6/44$ & $1\ (1)$ & $38.5\ (76.9)$ & $3.92$ \\
    $\SU(3)$ & ${\bf 6}_S + 7\times {\bf \overline 3}_F$ & C & $27$ & $12/33$ & $2\ (1)$ & $31.6\ (65.6)$ & $1.79$ \\
    $\SU(3)$ & ${\bf 15} + 2\times {\bf \overline 6}_S$ & C & $27$ & $30/33$ & $1\ (0)$ & $31.6\ (-)$ & $1.79$ \\
    $\SU(6)$ & ${\bf 15}_A + 2\times {\bf \overline 6}_F$ & C & $27$ & $6/66$ & $10\ (5)$ & $58.6\ (116.)$ & $6.44$ \\
    $E_6$ & ${\bf 27}$ & C & $27$ & $1/22$ & $21\ (11)$ & $102.\ (197.)$ & $13.8$\\
    $\SO(11)$ & ${\bf 32}_{\rm Spin}$ & PR & $32$ & $8/99$ & $1\ (1)$ & $83\ (163.)$ & $9.96$\\
    $\SO(12)$ & ${\bf 32}_{\rm Spin}$ & PR & $32$ & $8/110$ & $1\ (1)$ & $94\ (184.)$ & $11.9$\\
    $\SU(4)$ & ${\bf 20} + {\bf \overline{6}}_S + {\bf 4}_F$ & C & $34$ & $38/44$ & $1\ (0)$ & $44.8\ (-)$ & $3.10$
			\end{tabular}
	\caption{Models with massless families with less than 40 fermions. The multiplicities are computed for one family. For the indicated type, C means chiral (complex) while PR means pseudo-real. Values in parentheses correspond to the supersymmetric case, where allowed by asymptotic freedom. The $n_g^f$ column is set at maximum to 1 for PR representations as only one massless generation is possible in that case.} \label{tab:ng1}
\end{table}
Asymptotic freedom plays a crucial role in defining the theory degrees-of-freedom. Furthermore, if lost, it would deem the theory effective, as some couplings would diverge at a high scale. To search for the asymptotically free theories, one requires the first coefficient of the one-loop beta function to be negative. This implies
\begin{equation}
    b_{1-loop} = - \frac{11}{3} C_G \left( 1 - \sum_f n_g^f \xi_f \right)  < 0\,,
\end{equation}
with
\begin{equation}
    \xi_f = \frac{2}{11 C_G} \sum_{r\in f} T_r\,,
\end{equation}
where $T_r$ is the trace normalization of the representation $r$ and $C_G$ the quadratic Casimir of the adjoint.
This allows us to define the maximal multiplicity for complex families
\begin{equation}
    n_g^f < 1/\xi_f\,.
\end{equation}
For SUSY theories:
\begin{equation}
    b_{1-loop}^S = - 3 C_G \left( 1 - \frac{11}{6} \sum_f n_g^f \xi_f \right)\,,
\end{equation}
hence reducing the number of allowed families, $n_g^f < 6/(11\xi_f)$.

\begin{table}[t]
	\centering
	\begin{tabular}{l|c|cc|cc}
	Group & Family & $n_f$ & $\xi_f$ &  $f_{FE}$ & $f_a$ \\\hline
    $\SU(5)$ & ${\bf 10}_A + {\bf \overline{5}}_F$ & $15$ & $4/55$ & $63.4\ (129.)$ & $4.82$\\
    $\SO(10)$ & ${\bf 16}_{\rm Spin}$ & $16$ & $4/88$ & $87\ (174.)$ & $8.48$\\
    $\SU(6)$ & ${\bf 15}_A + 2\times {\bf \overline 6}_F$ & $27$ & $6/66$ & $106.\ (218.)$ & $7.27$ \\
    $\SU(4)$ & ${\bf 10}_S + 8 \times {\bf \overline{4}}_F$ & $42$ & $14/44$  & $125.\ (-)$ & $4.51$\\
    $E_6$ & ${\bf 27}$ & $27$ & $6/132$ & $149.\ (298.)$ & $14.7$\\
    $\SU(7)$ & ${\bf 21}_A + 3\times{\bf \overline 7}_F$ & $42$ & $8/77$ &  $158.\ (326.)$ & $10.2$\\
    $\SU(7)$ & ${\bf 35}_{A3} + 2\times{\bf \overline 7}_F$ & $49$ & $12/77$ & $177.\ (366.)$ & $10.5$\\
    $\SU(5)$ & ${\bf 15}_S + 9 \times {\bf \overline 5}_F$ & $60$ & $16/55$ &  $181.\ (-)$ & $6.88$\\
			\end{tabular}
	\caption{Models with $N_g = 3$ identical generations  and $f_{FE} < 200$. The models are listed with increasing free energy $f_{FE}$.  The $\SU(N)$ representations are the fundamental ($F$), the two-index symmetric ($S$), and the  two-index antisymmetric ($A$) and three-index ($A3$)  antisymmetric.} \label{tab:ng3}
\end{table}

\begin{table}[b]
	\centering
	\begin{tabular}{l|c|cc|cc}
	Group & 3$^{\rm rd}$ Family & $n_{f,3}$ & $\xi_{f,3}$ &  $f_{FE}$ & $f_a$ \\\hline
    $\SU(5)$ & ${\bf 10}_A + {\bf \overline{5}}_F$ & $15$ & $4/55$ & $63.4\ (129.)$ & $4.82$\\
    $\SU(5)$ & ${\bf 15}_S + 9\times {\bf \overline{5}}_F$ & $60$ & $16/55$ & $103.\ (214.)$ & $5.51$\\
    $\SU(5)$ & ${\bf 45} + 6\times {\bf \overline{5}}_F$ & $75$ & $30/55$ & $116.\ (-)$ & $5.74$\\
			\end{tabular}
	\caption{$\SU(5)$ models with 2 generations of ${\bf 10}_A + {\bf \overline{5}}_F$ and a third family of different type. We list all cases with $f_{FE} < 120$ for illustration.} \label{tab:ng3su5}
\end{table}

It is evident that minimizing either $f_{FE}$ or $f_a$ can only lead to the trivial minimal number of families, i.e. $N_g = 0$ (a pure Yang-Mills theory). So, for now, we will consider $N_g$ as an input and compare theories with the same number of families. In Table~\ref{tab:ng3} we list all the theories with $N_g = 3$ identical chiral families listed for increasing $f_{FE}$ (up to $f_{FE} < 200$). As explained above, they all correspond to complex representations. The model with minimal free energy is the $\SU(5)$ one of Georgi-Glashow \cite{Georgi:1974sy} with $f_{FE} = 63.4$, closely followed by $\SO(10)$ \cite{Fritzsch:1974nn} with $f_{FE} = 87$. Note also that the only model in the list that cannot contain the SM gauge symmetry is based on  $\SU(4)$. The same ranking is valid in the SUSY case, where the $\SU(4)$ model is excluded for loss of asymptotic freedom.
As expected, $f_a$ gives larger weights to the gauge part, so that the $\SU(5)$ model is only surpassed by $\SU(4)$, which has a very similar score.

For completeness, we also considered the case in which the three families consist of different representations. There are many such examples, with smaller free energy achieved when only one generation is different from the minimal case. We list the most minimal cases in Table~\ref{tab:ng3su5}: all these models, however, have $f_{FE}$ scores higher than the minimal $\SU(5)$ and $\SO(10)$ theories.

Our first finding, therefore, is that the free energy allows us to select the canonical $\SU(5)$ and $\SO(10)$ GUTs as the preferred models \emph{with three generations}. So far we have not discussed the scalar sector, which is crucial for breaking the GUT symmetry and generating appropriate Yukawa couplings.  In our framework, it would be ideal to require that the additional couplings are also asymptotically free, hence imposing non-trivial constraints on the model, which we leave for future investigation.

\section{Swampland of Gauge dualities}

The traditional approach to GUTs starts from the SM at low energies and aims at finding a minimal GUT that leads to the breaking of the unified gauge symmetry.
However, the gauge structure of the SM may be modified before the GUT scale is reached. For example, partial unification of leptons and quarks \cite{Pati:1974yy} could occur at energies as low as $10^{6}$~GeV. A complete classification of possible gauge extensions of the SM has been presented in \cite{Allanach:2021bfe}.

Another possibility is offered by gauge-gauge dualities, where one of the SM gauge groups emerges as the effective infra-red limit of a more fundamental ultra-violet theory. A concrete example was first provided for $\mathcal{N}=1$ SUSY $\SU(N)$ theories \cite{Seiberg:1994bz,Seiberg:1994pq}, and extended to non-SUSY theories in \cite{Mojaza:2011rw,Sannino:2011mr,Antipin:2011ny} due to the marginal role played by scalars. Such dualities can be applied straightforwardly to the $\SU(3)$ gauge of the SM \cite{Cacciapaglia:2024mfy} (for the SUSY case, see \cite{Maekawa:1995cz,Maekawa:1995ww}). An interesting feature of this ultra-violet completion is that the electroweak symmetry breaking is due to composite scalars at low energies and a multi-Higgs model naturally emerges, with features similar to scalar democracy \cite{Hill:2019ldq}. Furthermore, the number of colors in the dual theory is linked to the number of generations \cite{Sannino:2011mr} in the SM: at high energies, the dual gauge symmetry is $\SU(2N_g-3)$, where the duality is only possible for $N_g=3$ ($\SU(3)$), $N_g=4$ ($\SU(5)$) and $N_g=5$ ($\SU(7)$). Does the GTA allow us to select one high-energy theory out of these three, hence fixing the number of generations in the SM? 

\begin{table}[bt]
	\centering
	\begin{tabular}{l|ccc|}
	 & $\SU(X)$ & $\SU(2)_L$ & $\U(1)_Y$   \\\hline  
    $Q_L$ & $F$ & $2$ & $1/(2X)$ \\
    $U_R^c$ & $\overline{F}$ & $1$ & $-1/2-1/(2X)$  \\
    $D_R^c$ & $\overline{F}$ & $1$ & $1/2-1/(2X)$\\
    $l_L$ & $1$ & $2$ & $-1/2$\\
    $e_R^c$ & $1$ & $1$ & $1$\\ 
	\end{tabular}
	\caption{Fermion content of the high-energy dual SM, where $X = 2 N_g - 3$ and $N_g = 3,4,5$ is the number of allowed fermion generations. Hypercharge is normalized to give electric charge $Q = T_L^3 + Y$.} \label{tab:dual}
\end{table}

The fermionic matter content of the dual SM is listed in Table~\ref{tab:dual}, where $X = 2 N_g -3$. For $N_g=3$, this is exactly equivalent to a scalarless SM. The free energy considerations from Table~\ref{tab:ng3} already suggest that the $N_g=3$ case is to be preferred, as it can be embedded into the minimal-$f_{FE}$ $\SU(5)$ model with three generations. The straightforward extension to $N_g = 4,5$ generations would require to consider GUTs based on $\SU(7)$ and $\SU(9)$ respectively: however, no model in the GTA contains the chiral matter content in Table~\ref{tab:dual}. This is due to a special feature of $X=3$ \cite{Georgi:1980pd}, namely the fact that a two-index antisymmetric representation is equal to the anti-fundamental of $\SU(3)$, $A \equiv \overline{F}$. This feature allows embedding of $Q_L$, $T_R^c$ and $e_R^c$ in the same $\bf 10$ of $\SU(5)$. This embedding is not extendable to $\SU(7)$ nor $\SU(9)$.

Similar considerations apply to the $\SO(10)$ case, which could be extended to $\SO(14)$ and $\SO(18)$, respectively, for $N_g = 4,5$. The embedding of the $X=3$, $N_g=3$, case in $\SO(10)$ is based on the fact that $\SO(10) \supset \SO(6) \times \SO(4)$, where $\SO(6) \sim \SU(4)$ is equivalent to the fourth color gauge structure \cite{Pati:1974yy}. Also, $\SO(10) \supset \SU(5)$, hence containing the Georgi-Glashow model. Such considerations and embeddings cannot be extended to $\SO(14)$ and $\SO(18)$. Finally, $E_6 \supset \SU(5) \times \SU(2) \times \U(1)$ does not contain the fermion spectrum in Table~\ref{tab:dual}. Hence, we find that the dual SM with $4$ and $5$ generations are not in the GTA dryland. Note that the dual theory, which emerges from the GTA, does not contain any Yukawa couplings: they are generated by the duality itself at low energies \cite{Cacciapaglia:2024mfy}. Hence, a realistic dual GUT would only require a single scalar field at high scale for the breaking of the gauge symmetry down to the dual SM one.

\section{Which gauge extension?}

The GTA dryland can also be used to constrain possible extensions of the SM gauge group at low energies.
This is common practice in model building: for instance, in replacing the Higgs sector with a strongly coupled gauge interaction \cite{Cacciapaglia:2020kgq} or in including a dark sector to explain dark matter \cite{Antipin:2015xia,Contino:2020god}. In both cases, the extended gauge group has a factorized form $\mathcal{G}_{\rm SM} \times \mathcal{G}_{\rm BSM}$. What extensions of this type are in the dryland of the GTA?

As an example, we analyzed models consisting of $\SU(5) \times \SU(N) \subset \SU(N+5)$, where the SM is contained in the $\SU(5)$ gauge part. The minimal requirement is $N\geq 2$ in order to obtain a non-abelian BSM sector that confines at low energies. The $\SU(N+5)$ chiral families should contain three generations of Georgi-Glashow multiplets, i.e. $\bf{10} + \bf{\overline 5}$, plus states charged under $\SU(N)$. After analyzing all chiral families for $N\geq 2$, we found a unique viable model. It is based on $\SU(8)$, with one family of
\begin{equation}
    {\bf \overline{56}}_{A3} + 2 \times {\bf 28}_A + 3 \times {\bf \overline{8}}_F\,, 
\end{equation}
which has $n_f = 136$ and $\xi_f = 30/88$. Under $\SU(8) \to \SU(5) \times \SU(3) \times \U(1)$, each representation decomposes as:
\begin{eqnarray}
    {\bf \overline{56}} &\to & (10,1)_{-9} \oplus (\overline{10},\overline{3})_{-1} \oplus (\overline{5}, 3)_{7} \oplus (1,1)_{15}\,, \\
    {\bf 28} &\to & (10, 1)_{6} \oplus (5,3)_{-2} \oplus (1,\overline{3})_{-10}\,, \\
    {\bf \overline{8}} &\to & (\overline{5}, 1)_{-3} \oplus (1,\overline{3})_{5}\,.
\end{eqnarray}
Taking into account the multiplicities in the $\SU(8)$ family, we note the presence of three $\SU(5)$ families plus one singlet, and various fields charged under $\SU(3)$:
\begin{eqnarray}
    F\,  &\Rightarrow & (\overline{5}, 3)_{7} + 2\times (5, 3)_{-2}\,, \\
    \overline{F}\, &\Rightarrow & (\overline{10}, \overline{3})_{1} + 2\times (1,\overline{3})_{-10} + 3\times (1,\overline{3})_{5}\,.
\end{eqnarray}
If this $\SU(3)$ gauge interaction generated confinement and chiral symmetry breaking at low energies, the mesons will contain scalars that transform as ${\bf 5}$ and ${\bf 45}$ of $\SU(5)$, where both contain a Higgs component. If a vacuum expectation value is generated for such mesons, they could be identified with the source of electroweak symmetry breaking, and hence play the role of the SM Higgs. Naturally, the breaking of $\SU(8) \to \SU(5)\times \SU(3) \to \mathcal{G}_{\rm SM} \times \SU(3)$ must be obtained via (heavy) elementary scalars, while the confinement of $\SU(3)$ must occur not far from the electroweak scale. Baryons that transform as one family of SM fermions are also generated, and hence they could play the role of partial compositeness partners for the third generation of quarks (and leptons), as in \cite{Vecchi:2015fma}.

\section{Outlook}

In this article, we proposed an atlas of asymptotically free gauge theories, based on a simple unified gauge factor, with massless families of fermions. The free energy at high temperatures can be used as a counter for the degrees of freedom of each theory, thus allowing us to chart the atlas. By requiring three families of fermions, the free energy naturally selects the $\SU(5)$ Georgi-Glashow GUT as the most minimal theory, closely followed by the canonical $\SO(10)$ GUT. Hence, the SM may have a natural position within the GTA. The atlas can also be used to check if gauge extensions of the SM allow for a grand-unified extension at high energies. We illustrated this in two examples, one based on a gauge-gauge duality on the chromomagnetic interactions, and the other based on an $\SU(N)$ extension of the standard gauge symmetry. In the former case, the GTA yields a strong preference for three families.

We have not considered yet scalar fields, as their mass cannot be prevented by any symmetry. Including spontaneous symmetry breaking of the gauge symmetry and generation of Yukawa couplings could imprint further constraints on the atlas, providing a phenomenological compass to navigate us towards the optimal high-energy theory. In our analysis, asymptotic freedom plays a crucial role in counting the degrees of freedom of each theory. 

\paragraph{Acknowledgements}

\noindent
The work of F.S. is partially supported by the Carlsberg Foundation, grant CF22-0922.

\vspace{0.3cm}

\bibliography{biblio}

\widetext
\newpage

\section*{Appendix}

\renewcommand\thetable{S--\arabic{table}}
\setcounter{table}{0}
\renewcommand\thepage{\roman{page}}
\setcounter{page}{1}

\subsection*{Sp(2N) pseudo-real families and the Witten anomalies}

The well-known $\SU(2)$ Witten anomaly~\cite{Witten:1982fp} admits a natural generalization to the symplectic groups $\Sp(2n)$, as Witten’s original argument relies on the non-triviality of the fourth homotopy group $\pi_4$. The family of Lie groups $\Sp(2n)$ is the only one for which $\pi_4 = \mathbb{Z}_2 \ne 0$. Furthermore, symplectic groups contain pseudo-real irreducible representations, which can be used to define theories within the GTA. 

In this context, the candidate carriers of the anomaly are Weyl fermions in pseudo-real representations. The group $\Sp(2n)$ admits only real or pseudo-real representations, due to the existence of an invariant symplectic form. Weyl fermions in real representations can be given gauge-invariant mass terms, allowing for a Pauli–Villars regularization scheme and thus avoiding any anomaly. In contrast, pseudo-real representations do not admit such mass terms.

\begin{table}[h]
	\centering
	\begin{tabular}{l|l|c|c|c|c}
	Name & Dynkin rep. & Dimension  & $2T_r$  & Anomaly? & AF  \\\hline
$\bf F$ & $(1,\dots)$ & $2n$ & $1$ & Yes &  Always \\
$\bf A3$ & $(0,0,1,\dots)$ & $\frac{2n}{3} (2n+1)(n-2)$ & $(n-1)(2n-3)-1$ & $n$ odd & $3\leq n\leq8$  \\
$\bf A5$ & $(0,0,0,0,1,\dots)$ & $\frac{n}{15} (n-4)(n-1)(4n^2-1)$ & $\frac{1}{6} (n-1)(n-4)(2n-3)(2n-1)$ & No & $n=5$  \\
$\bf S3$ & $(3,\dots)$ & $\frac{2n}{3} (2n+1)(n+1)$ & $(n+1)(2n+3)$ & $n$ even & $1\leq n\leq3$  \\ 
$\bf R11$ & $(1,1,\dots)$ & $\frac{8}{3} n(n^2-1)$ & $4(n^2-1)$ & Never & $2\leq n \leq 3$ \\
\end{tabular}
	\caption{List of $\Sp(2n)$ irreducible pseudo-real representations that respect asymptotic freedom. They are labeled by their Dynkin label, dimension and Dynkin index $T_r$. Conditions on their contribution to the Witten anomaly and asymptotic freedom are provided in the last two columns.} \label{tab:sp2nirreps}
\end{table}

The pseudo-real representations of $\Sp(2n)$ are precisely those that appear in the odd tensor powers of the defining representation $\mathbf{F} \equiv \mathbf{2n}$~\cite{okubo_further_1989-1}. In Table~\ref{tab:sp2nirreps} we list all the pseudo-real representations that leave the gauge coupling asymptotically free (AF). We use the Dynkin labels, in the second column, to identify the irreducible representations. Among these, only those with a \emph{half-integer Dynkin index} contribute non-trivially to the Witten anomaly, assuming the normalization $\operatorname{T}(\mathbf{2n}) = 1/2$~\cite{Okubo:1989}. Besides the fundamental, which always carries the Witten anomaly \cite{Witten:1982fp}, the list contains three-index representations. For every $\Sp(2n)$ group, either the symmetric or the anti-symmetric carry the anomaly depending on $n$ being even (symmetric) or odd (anti-symmetric). Instead, the mixed representation $(1,1,\dots)$ always has integer Dunkin index. Finally, the five-index anti-symmetric is only relevant for $\Sp(10)$, the first group for which it is defined, as it looses AF for larger groups.

In the presence of multiple fermion species, the anomaly is given by the sum of their Dynkin indices~\cite{Wang:2018qoy}. Accordingly, the anomaly can be canceled by including an even number of representations that are individually anomalous. If those representations are different, mass terms remain forbidden, hence allowing to define anomaly-free pseudo-real families.
Taking these considerations into account—along with asymptotic freedom descending from the GTA conditions—we construct Table~\ref{tab:sp2n}.

\begin{table}[h]
	\centering
	\begin{tabular}{l|c|ccc|cc|c}
	Group & Family  & $n_f$  & $\xi_f$ & $n_g^f \leq$ & $f_{FE}$  & $f_a$ & WWW anomaly? \\\hline
    $\Sp(2)$ & ${\bf 4}_{S3}$ & $4$ & $14/22$ & $1\ (0)$ & $6.5\ (-)$ & $0.58$ & Yes\\
    $\Sp(4)$ & ${\bf 16}_{R11}$ & $16$ & $12/33$ & $1\ (1)$ & $24\ (48.7)$ & $1.97$ & No\\
    $\Sp(4)$ & ${\bf 4}_F + {\bf 20}_{S3}$  & $24$ & $22/33$ & $1\ (0)$ & $31\ (-)$ & $2.09$ & Yes\\ 
    $\Sp(6)$ & ${\bf 6}_F + {\bf 14}_{A3}$  & $20$ & $6/44$ & $1\ (1)$ & $38.5\ (76.9)$ & $3.92$ & No \\ 
    $\Sp(6)$ & ${\bf 56}_{S3}$  & $56$ & $36/44$ & $1\ (0)$ & $70\ (-)$ & $4.47$ & Yes\\
    $\Sp(6)$ & ${\bf 64}_{R11}$  & $64$ & $32/44$ & $1\ (0)$ & $77\ (-)$ & $4.59$ & No\\
    $\Sp(8)$ & ${\bf 48}_{A3}$  & $48$ & $14/55$ & $1\ (1)$ & $78\ (158.)$ & $6.93$ & No\\
    $\Sp(10)$ & ${\bf 10}_{F} + {\bf 110}_{A3}$  & $120$ & $28/66$ & $1\ (1)$ & $160\ (328.)$ & $11.3$ & No\\
    $\Sp(10)$ & ${\bf 132}_{A5}$  & $132$ & $42/66$ & $1\ (0)$ & $170.\ (328.)$ & $11.5$ & No \\
    $\Sp(12)$ & ${\bf 208}_{A3}$  & $208$ & $44/88$ & $1\ (1)$ & $260\ (536.)$ & $16.6$ & No\\
    $\Sp(14)$ & ${\bf 14}_{F} + {\bf 350}_{A3}$  & $364$ & $66/99$ & $1\ (-)$ & $432.\ (-)$ & $23.6$ & No\\
    $\Sp(16)$ & ${\bf 544}_{A3}$  & $544$ & $90/110$ & $1\ (-)$ & $612\ (-)$ & $31.7$ & No\\
            \end{tabular}
	\caption{Complete list of GTA models based on $\Sp(2N)$ gauge theories with pseudo-real families free of the Witten anomaly. The last column indicates if the model suffers from the Wang-Wen-Witten anomaly \cite{Wang:2018qoy}, associated with the three-index symmetric representation.} \label{tab:sp2n}
\end{table}

Some of these models can be excluded by a second $\SU(2)$ anomaly, discovered in \cite{Wang:2018qoy}. This anomaly typically applies to three-index symmetric representations. However, it is based on the assumption that all fermion fields are in odd tensor powers of the defining representation (pseudo-real) while all bosons are in even tensors (real). This hypothesis could be invalidated by the presence of massive states, such as fermions in real representations. Nevertheless, in the main text we have excluded theories that carry this anomaly.

\subsection*{Chiral SU(N) families}

Complex (chiral) families are constructed by complex representations, of which $\SU(n)$ groups are rich. The only two constraints are given by the cancellation of the gauge anomalies and by asymptotic freedom. A list of all complex and AF representations is presented in Table~\ref{tab:sunirreps}, adapted from \cite{Eichten:1981mu}. The AF limit on $n$ in the next-to-last column is derived by counting the contribution of the representation alone. In practice, when constructing a chiral family, additional representations must be added to cancel the gauge anomaly. This does not affect the representations up to two tensor indices, ${\bf F}$, ${\bf A2}$ and ${\bf S2}$. For higher rank representation, however, this effects is crucial. To estimate a conservative range, we use the fact that ${\bf S2}$ is the most efficient representation, in the sense that it provides the smallest Dynkin index per unit of anomaly, hence we used the inequality
\begin{equation}
    2 T_r + \left|\frac{\mathcal{A}_r}{\mathcal{A}_{S2}}\right|\ 2 T_{S2}  < 11 n
\end{equation}
to obtain the largest $n$ allowing for a AF chiral family containing the representation $r$. The result is listed in the last column of Table~\ref{tab:sunirreps}.

\begin{table}[h]
	\centering
	\begin{tabular}{l|l|c|c|c|c|c}
	Name & Dynkin rep. & Dimension  & $2T_r$  & Anomaly $\mathcal{A}_r$ & AF & AF in family \\\hline
${\bf F}$ & $(1,\dots)$ & $n$ & $1$ & $1$ &  $n\geq 3$ & All\\
$\bf A2$ & $(0,1,\dots)$ & $\frac{1}{2} n (n-1)$ & $(n-2)$ & $n-4$ & $ n\geq 5$ & All \\
$\bf S2$ & $(2,\dots)$ & $\frac{1}{2} n (n+1)$ & $(n+2)$ & $n+4$ & $n\geq 3$ & All \\
$\bf A3$ & $(0,0,1,\dots)$ & $\frac{1}{6} n (n-1) (n-2)$ & $\frac{1}{2} (n-2)(n-3)$ & $\frac{1}{2} (n-3)(n-6)$ & $ 7\leq n\leq 26$ & $7\leq n\leq 17$ \\
$\bf A4$ & $(0,0,0,1,\dots)$ & $\frac{1}{24} n (n-1) (n-2) (n-3)$ & $\frac{1}{6} (n-2)(n-3)(n-4)$ & $\frac{1}{6} (n-3)(n-4)(n-8)$ & $ 9\leq n\leq 12$ & $9\leq n\leq 11$ \\
$\bf S3$ & $(3,\dots)$ & $\frac{1}{6} n (n+1) (n+2)$ & $\frac{1}{2} (n+2)(n+3)$ & $\frac{1}{2} (n+3)(n+6)$ & $ 3\leq n\leq 16$ & Never \\
$\bf R11$ & $(1,1,\dots)$ & $\frac{1}{3} n (n^2-1)$ & $ (n^2-3)$ & $ (n^2-9)$ & $ 4\leq n\leq 11$ & $4\leq n\leq 6$ \\
$\bf R22$ & $(0,2,\dots)$ & $\frac{1}{12} n^2 (n^2-1)$ & $\frac{1}{3} n(n^2-4)$ & $\frac{1}{3} n (n^2-16)$ & $ 5\leq n\leq 6$ & $n = 5$ \\
$\bf Ra$ & $(1,0,\dots,1,0)$ & $\frac{1}{2} n (n+1)(n-2)$ & $\frac{1}{2} (n-2)(3n+1)$ & $\frac{1}{2} (-n^2 +7n+2)$ & $ 5\leq n\leq 9$ & $5\leq n\leq 8$ \\
$\bf R13$ & $(1,0,1,\dots)$ & $\frac{1}{8} n (n^2-1)(n-2)$ & $\frac{1}{2} (n-2)(n^2-n-4)$ & $\frac{1}{2} (n-4) (n^2 -n-8)$ & $ n= 6$ & Never \\
$\bf Rb$ & $(2,\dots,1)$ & $\frac{1}{2} n (n-1)(n+2)$ & $\frac{1}{2} (n+2)(3n-1)$ & $\frac{1}{2} (n^2 +7n-2)$ & $ 3\leq n\leq 5$ & $n = 3$ \\
\end{tabular}
	\caption{List of $\SU(n)$ irreducible complex representations that respect asymptotic freedom. They are labeled by their Dynkin label, dimension and Dynkin index $T_r$. In the last three columns, we indicate the contribution to the gauge anomaly, and the conditions for asymptotic freedom. Note that ${\bf Ra} \equiv {\bf R13}$ for $\SU(5)$.} \label{tab:sunirreps}
\end{table}

The explicit construction of the AF chiral families crucially depends on the rank of the $\SU(n)$ group and, consequently, on the number of involved representations. As the representations of rank less than two, $\bf F$, $\bf A2$ and $\bf S2$, are always present for every $n$, it is first convenient to build all irreducible families made of these three. The result crucially depends on $n$. Nevertheless, one can identify three families that are always present for every $n$:
\begin{eqnarray}
    {\bf A2} + (n-4) \ {\bf \overline{F}} \quad \mbox{for}\;\; n\geq 5\,, \\
    {\bf S2} + (n+4)\ {\bf \overline{F}} \quad \mbox{for}\;\; n\geq 3\,, \\
    {\bf S2} + {\bf \overline{A2}} + 8\ {\bf \overline{F}} \quad \mbox{for}\;\; n\geq 5\,.
\end{eqnarray}
The first corresponds to a generalization of the Georgi-Glashow model \cite{Georgi:1974sy}, while the second class was first identified by Bars and Yankielowicz \cite{Bars:1981se}.
For every $n$, other irreducible combinations are also allowed, which crucially depend on $n$.
On the other hand, for $n\geq 18$, only these three representations are allowed by AF, leading to classifiable families.

In the following, we will present the complete list of families for all $\SU(n)$ group, starting from the smallest.

\subsection{SU(3) and SU(4)}

The $\SU(3)$ and $\SU(4)$ are the smallest groups with complex representations, and they are characterized by the absence of the ${\bf A2}$ representation (which is equivalent to the $\bf \overline{F}$ for $\SU(3)$ and real for $\SU(4)$).
For both, only three representations allow to build chiral families, c.f.  Tables~\ref{tab:su3} and \ref{tab:su4}, left.

There are only three irreducible anomaly-free combinations that preserve asymptotic freedom in $\SU(3)$, and they are listed in the Table~\ref{tab:su3} right. For $\SU(4)$, we found four combinations, as listed in the Table~\ref{tab:su4} right.

\begin{table}[h]
	\centering
	\begin{tabular}{l|c|c|c|}
	 Representation  & $d_r$  & $2T_r$ & $\mathcal{A}_r$   \\\hline
${\bf F}$ & $3$ & $1$ & $1$ \\ 
${\bf S2}$ & $6$ & $5$ & $7$ \\
${\bf Rb}$ & $15$ & $20$ & $14$ \\
\end{tabular}
\hspace{1cm}
\begin{tabular}{l|c|c|c|}
	 $\SU(3)$ family  & $n_f$  & $33\ \xi_f$ & $n_g^f \leq$   \\\hline
${\bf 6}_{S2} + 7\ {\bf \overline{3}}_F$ & $27$ & $12$ & $2\ (1)$ \\ 
${\bf 15}_{Rb}+ {\bf \overline{6}}_{S2} + 7\ {\bf \overline{3}}_F$ & $42$ & $32$ & $1\ (0)$ \\
${\bf 15}_{Rb}+ 2\ {\bf \overline{6}}_{S2}$ & $27$ & $30$ & $1\ (0)$ \\
\end{tabular}
	\caption{Complete list of relevant representations (left table) and chiral families (right table) for $\SU(3)$. In the left table, the representations are labeled by their dimension. The maximum number of generations is shown in the last column, with the supersymmetric case in parenthesis.} \label{tab:su3}
\end{table}

\begin{table}[h]
	\centering
	\begin{tabular}{l|c|c|c|}
	 Representation  & $d_r$  & $2T_r$ & $\mathcal{A}_r$   \\\hline
${\bf F}$ & $4$ & $1$ & $1$ \\ 
${\bf S2}$ & $10$ & $6$ & $8$ \\
${\bf R11}$ & $20$ & $13$ & $7$ \\
\end{tabular}
\hspace{1cm}
\begin{tabular}{l|c|c|c|}
	 $\SU(4)$ family  & $n_f$  & $44\ \xi_f$ & $n_g^f \leq$   \\\hline
${\bf 10}_{S2} + 8\ {\bf \overline{4}}_F$ & $42$ & $14$ & $3\ (1)$ \\ 
${\bf 20}_{R11} + 7\ {\bf \overline{4}}_F$ & $48$ & $20$ & $2\ (1)$ \\
${\bf 20}_{R11}+ {\bf \overline{10}}_{S2} + {\bf 4}_F$ & $34$ & $20$ & $2\ (1)$ \\
2\ ${\bf 20}_{R11} + {\bf \overline{10}}_{S2} + 6\ {\bf \overline{4}}_F$ & $74$ & $38$ & $1\ (0)$ \\
\end{tabular}
	\caption{Complete list of relevant representations (left table) and chiral families (right table) for $\SU(4)$. In the left table, the representations are labeled by their dimension. The maximum number of generations is shown in the last column, with the supersymmetric case in parenthesis.} \label{tab:su4}
\end{table}

\subsection{SU(5)}

The group $\SU(5)$ is the smallest one that features a complex non-trivial $\bf A2$ representation. In total, there are six representation that can be used to build chiral families, c.f. Table~\ref{tab:su5} left.
The general strategy for $\SU(n)$ groups with $n\geq 5$ is to first construct all irreducible families made of the three lowest-rank representations: $\bf F$, $\bf A2$ and $\bf S2$. The number and structure of the independent families depend crucially on the contribution to the anomaly of the two rank-two representations. For $\SU(5)$, as $\mathcal{A}_{A2} = \mathcal{A}_{F} = 1$, the fundamental and anti-symmetric are interchangeable. This implies that, besides the Georgi-Glashow and Bars-Yankielowicz families, one can build a class of families made of ${\bf S2} + x\ {\bf \overline{A2}} + (9-x)\ {\bf \overline{F}}$, as listed in the top block of Table~\ref{tab:su5} right.


The second step consists in ranking the remaining higher-order representations by increasing $\mathcal{A}_r$, c.f. lower block of Table~\ref{tab:su5} left. The families are then constructed iteratively, by first using the $\bf \overline{F}$ to cancel the anomaly, and then substituting the needed number of fundamentals with the other representations. For instance, taking the $\bf Ra$ representation, with $\mathcal{A}_{Ra} = 6$, we first build ${\bf Ra} + 6\ {\bf \overline{F}}$; then one can substitute ${\bf F} \to {\bf A2}$ to construct the class ${\bf Ra} + x\ {\bf \overline{A2}} + (6-x)\ {\bf \overline{F}}$. Then one substitutes $9\ {\bf F} \to {\bf S2}$: as $\mathcal{A}_{Ra} < \mathcal{A}_{S2}$, the anomaly mismatch can be compensated as ${\bf Ra} + {\bf \overline{S2}} + 3\ {\bf F}$. Finally, the substitution ${\bf F} \to {\bf A2}$ is performed again. This leads to the families in the second block of Table~\ref{tab:su5} right. This step can be repeated for the other two representations, $\bf R22$ and $\bf R11$. Finally, one can consider combinations a starting point, where the only usable one turns out to be ${\bf R11} + {\bf \overline{R22}}$. The complete list of $52$ independent families is reported in Table~\ref{tab:su5} right.

\begin{table}[h]
	\centering
    	\begin{tabular}{l|c|c|c|}
	 Representation  & $d_r$  & $2T_r$ & $\mathcal{A}_r$   \\\hline
${\bf F}$ & $5$ & $1$ & $1$ \\ 
${\bf A2}$ & $10$ & $3$ & $1$ \\
${\bf S2}$ & $15$ & $7$ & $9$ \\ \hline
${\bf Ra}$ & $45$ & $24$ & $6$ \\
${\bf R22}$ & $50$ & $35$ & $15$ \\
${\bf R11}$ & $40$ & $22$ & $16$ \\
\end{tabular}
\hspace{0.5cm}
\begin{tabular}{l|c|c|c|c|}
	 $\SU(5)$ family  & $n_f$  & $55\ \xi_f$ & $n_g^f \leq$ & condition  \\\hline
${\bf 10}_{A2} +{\bf \overline{5}}_F$ & $15$ & $4$ & $13\ (7)$ & \\ 
${\bf 15}_{S2} + 9\ {\bf \overline{5}}_F$ & $60$ & $16$ & $3\ (1)$ & \\
${\bf 15}_{S2}+ {\bf \overline{10}}_{A2} + 8\ {\bf \overline{5}}_F$ & $65$ & $18$ & $3\ (1)$ & \\
${\bf 15}_{S2}+x\ {\bf \overline{10}}_{A2} + (9-x)\ {\bf \overline{5}}_F$ & $60+5x$ & $(16+2x)$ & $2\ (1)$ & $2\leq x\leq 5$\\
${\bf 15}_{S2}+x\ {\bf \overline{10}}_{A2} + (9-x)\ {\bf \overline{5}}_F$ & $60+5x$ & $(16+2x)$ & $1\ (1)$ & $x=6$\\
${\bf 15}_{S2}+x\ {\bf \overline{10}}_{A2} + (9-x)\ {\bf \overline{5}}_F$ & $60+5x$ & $(16+2x)$ & $1\ (0)$ & $7\leq x\leq 9$\\\hline
${\bf 45}_{Ra} + x\ {\bf \overline{10}}_{A2} +(6-x)\ {\bf \overline{5}}_F$ & $75+5x$ & $(30+2x)$ & $1\ (0)$ & $0\leq x\leq 6$\\
${\bf 45}_{Ra} + {\bf \overline{15}}_{S2} + x\ {\bf 10}_{A2} + (3-x)\ {\bf 5}_F$ & $75+5x$ & $(34+2x)$ & $1\ (0)$ & $0\leq x\leq 3$\\\hline
${\bf 50}_{R22} + x\ {\bf \overline{10}}_{A2}+ (15-x)\ {\bf \overline{5}}_F$ & $125+5x$ & $(50+2x)$ & $1\ (0)$ & $0\leq x\leq 2$\\
${\bf 50}_{R22} + {\bf \overline{15}}_{S2}+ x\ {\bf \overline{10}}_{A2} + (6-x)\ {\bf \overline{5}}_F$ & $95+5x$ & $(48+2x)$ & $1\ (0)$ & $0\leq x\leq 3$\\
${\bf 50}_{R22} + 2\ {\bf \overline{15}}_{S2}+ 3\ {\bf 5}_F$ & $95$ & $52$ & $1\ (0)$ & \\
${\bf 50}_{R22} + 2\ {\bf \overline{15}}_{S2}+ {\bf 10}_{A2} + 2\ {\bf 5}_F$ & $100$ & $54$ & $1\ (0)$ & \\\hline
${\bf 40}_{R11} + x\ {\bf \overline{10}}_{A2} + (16-x)\ {\bf \overline{5}}_F$ & $120+5x$ & $(38+2x)$ & $1\ (0)$ & $0\leq x\leq 8$\\
${\bf 40}_{R11} + {\bf \overline{15}}_{S2}+ x\ {\bf \overline{10}}_{A2} + (7-x)\ {\bf \overline{5}}_F$ & $90+5x$ & $(36+2x)$ & $1\ (0)$ & $0\leq x\leq 7$\\
${\bf 40}_{R11} + 2\ {\bf \overline{15}}_{S2}+ 2\ {\bf 5}_F$ & $80$ & $38$ & $1\ (0)$ & \\
${\bf 40}_{R11} + 2\ {\bf \overline{15}}_{S2}+ {\bf 10}_{A2} + {\bf 5}_F$ & $85$ & $40$ & $1\ (0)$ & \\
${\bf 40}_{R11} + 2\ {\bf \overline{15}}_{S2}+ 2\ {\bf 10}_{A2}$ & $90$ & $42$ & $1\ (0)$ & \\
${\bf 40}_{R11} + {\bf \overline{45}}_{Ra}+ {\bf \overline{15}}_{S2} + {\bf \overline{5}}_F$ & $105$ & $54$ & $1\ (0)$ & \\
\end{tabular}
	\caption{Complete list of  chiral families (right table) for $\SU(5)$, with representations labeled by their dimension.} \label{tab:su5}
\end{table}

\subsection{Large n families}

\begin{table}[h]
	\centering
	\begin{tabular}{l|c|c|c|c|}
	 Family  & $n_f$  & $11 n \xi_f$ & $n_g^f <$ &  \\\hline
${\bf A2} + (n-4)\ {\bf \overline{F}}$ & $\frac{3}{2} n (n-3)$ & $2(n-3)$ & $\frac{11 n}{2(n-3)}$ & $n\geq 5$ \\ 
${\bf S2} + (n+4)\ {\bf \overline{F}}$ & $\frac{3}{2} n (n+3)$ & $2(n+3)$ & $\frac{11 n}{2(n+3)}$ & $n\geq 3$ \\
${\bf S2} + {\bf \overline{A2}} + 8\ {\bf \overline{F}}$ & $n(n+8)$ & $2(n+4)$ & $\frac{11n}{2(n+4)}$ & $n\geq 5$ \\  \hline
${\bf S2} + 2\ {\bf \overline{A2}} + (n-12)\ {\bf F}$ & $\frac{5}{2} n(n-5)$ & $2(2n+7)$ & $\frac{11n}{2(2n+7)}$ & $n\geq 16$ \\
$2\ {\bf S2} + 3\ {\bf \overline{A2}} + (20-n)\ {\bf \overline{F}}$ & $\frac{3}{2} n(n+13)$ & $2(2n+9)$ & $\frac{11n}{2(2n+9)}$ & $ 16\leq n \leq 20$ \\
$2\ {\bf S2} + 3\ {\bf \overline{A2}} + (n-20)\ {\bf F}$ & $\frac{1}{2} n(7n-41)$ & $2(3n-11)$ & $\frac{11n}{2(3n-11)}$ & $ n \geq 20$ \\
$3\ {\bf S2} + 4\ {\bf \overline{A2}} + (28-n)\ {\bf \overline{F}}$ & $\frac{5}{2} n(n+11)$ & $2(3n+13)$ & $\frac{11n}{2(3n+12)}$ & $ 21\leq n \leq 28$ \\
$3\ {\bf S2} + 4\ {\bf \overline{A2}} + (n-28)\ {\bf F}$ & $\frac{3}{2} n(3n-19)$ & $2(4n-15)$ & $\frac{11n}{2(4n-15)}$ & $ n \geq 28$ \\
$4\ {\bf S2} + 5\ {\bf \overline{A2}} + (36-n)\ {\bf \overline{F}}$ & $\frac{1}{2} n(7n+71)$ & $2(4n+17)$ & $\frac{11n}{2(4n+17)}$ & $ 29\leq n \leq 36$ \\
$4\ {\bf S2} + 5\ {\bf \overline{A2}} + (n-36)\ {\bf F}$ & $\frac{1}{2} n(11n-73)$ & $2(5n-19)$ & $\frac{11n}{2(5n-19)}$ & $ n \geq 36$ \\
$3\ {\bf S2} + 5\ {\bf \overline{A2}} + (2n-32)\ {\bf F}$ & $3 n(2n-11)$ & $2(5n-18)$ & $\frac{11n}{2(5n-18)}$ & $ 16\leq n \leq 19$ \\
\end{tabular}
	\caption{Complete list of chiral $\SU(n)$ families for $n\geq 16$ (for $n=16,17$, additional family exists with one $A_3$), see Tables \ref{tab:su16} and \ref{tab:su17}.} \label{tab:largeNsun}
\end{table}

For large $n$, namely $n \geq 18$, only three representations can be used to define chiral families: $\bf F$, $\bf A2$ and $\bf S2$. This allows us to classify all possible irreducible and anomaly-free combinations, as listed in Table~\ref{tab:largeNsun}.
In fact, this pattern also continues for $n=16,17$, with the only difference that  additional families can be added, containing one $\bf A3$ representation (see Tables \ref{tab:su16} and \ref{tab:su17}).

For $n\leq 15$, the patterns change, and more representations must be considered, hence the construction of chiral families must proceed case by case.

\subsection{SU(6) to SU(17)}

We complete the classification by presenting the complete list of chiral families for $\SU(6)$ in Table~\ref{tab:su6}, $\SU(7)$ in Table~\ref{tab:su7}, $\SU(8)$ in Table~\ref{tab:su8}, $\SU(9)$ in Table~\ref{tab:su9}, $\SU(10)$ in Table~\ref{tab:su10}, $\SU(11)$ in Table~\ref{tab:su11}, $\SU(12)$ in Table~\ref{tab:su12}, $\SU(13)$ in Table~\ref{tab:su13}, $\SU(14)$ in Table~\ref{tab:su14}, $\SU(15)$ in Table~\ref{tab:su15}, $\SU(16)$ in Table~\ref{tab:su16} and $\SU(17)$ in Table~\ref{tab:su17}.

\begin{table}[h]
	\centering
    	\begin{tabular}{l|c|c|c|}
	 Representation  & $d_r$  & $2T_r$ & $\mathcal{A}_r$   \\\hline
${\bf F}$ & $6$ & $1$ & $1$ \\ 
${\bf A2}$ & $15$ & $4$ & $2$ \\
${\bf S2}$ & $21$ & $8$ & $10$ \\ \hline
${\bf Ra}$ & $84$ & $38$ & $4$ \\
${\bf R11}$ & $70$ & $33$ & $27$ \\
\end{tabular}
\hspace{0.5cm}
\begin{tabular}{l|c|c|c|c|}
	 $\SU(6)$ family  & $n_f$  & $66\ \xi_f$ & $n_g^f \leq$ & condition  \\\hline
${\bf 15}_{A2} +2\ {\bf \overline{6}}_F$ & $27$ & $6$ & $10\ (5)$ & \\ 
${\bf 21}_{S2}+ x\ {\bf \overline{15}}_{A2} + (10-2x)\ {\bf \overline{6}}_F$ & $81+3x$ & $(18+2x)$ & $3\ (1)$ & $0\leq x \leq 1$\\
${\bf 21}_{S2}+ x\ {\bf \overline{15}}_{A2} + (10-2x)\ {\bf \overline{6}}_F$ & $81+3x$ & $(18+2x)$ & $2\ (1)$ & $2\leq x\leq 5$\\
\hline
${\bf 84}_{Ra} + x\ {\bf \overline{15}}_{A2}+ (4-2x)\ {\bf \overline{6}}_F$ & $108+3x$ & $(42+2x)$ & $1\ (0)$ & $0\leq x \leq2$\\
${\bf 84}_{Ra} + {\bf \overline{21}}_{S2}+ x\ {\bf 15}_{A2} + (6-2x)\ {\bf 6}_F$ & $141+3x$ & $(52+2x)$ & $1\ (0)$ & $0\leq x\leq 3$\\\hline
${\bf 70}_{R11} + x\ {\bf \overline{15}}_{A2}+ (27-2x)\ {\bf \overline{6}}_{F}$ & $232+3x$ & $(60+2x)$ & $1\ (0)$ & $0\leq x\leq 2$\\
${\bf 70}_{R11} + {\bf \overline{21}}_{S2} + x\ {\bf \overline{15}}_{A2} + (17-2x)\ {\bf \overline{6}}_{F}$ & $193+3x$ & $(58+2x)$ & $1\ (0)$ & $0\leq x\leq 3$\\
${\bf 70}_{R11} + 2\ {\bf \overline{21}}_{S2} + x\ {\bf \overline{15}}_{A2} +(7-2x)\  {\bf 6}_{F}$ & $154+3x$ & $(56+2x)$ & $1\ (0)$ & $0\leq x\leq 3$\\
${\bf 70}_{R11} +3\ {\bf \overline{21}}_{S2} +  3\ {\bf 6}_{F}$ & $151$ & $60$ & $1\ (0)$ & \\
${\bf 70}_{R11} +3\ {\bf \overline{21}}_{S2} + {\bf 15}_{A2} + {\bf 6}_{F}$ & $154$ & $62$ & $1\ (0)$ & \\
\end{tabular}
	\caption{Complete list of  chiral families (right table) for $\SU(6)$, with representations labeled by their dimension.} \label{tab:su6}
\end{table}

\begin{table}[h]
	\centering
    	\begin{tabular}{l|c|c|c|}
	 Representation  & $d_r$  & $2T_r$ & $\mathcal{A}_r$   \\\hline
${\bf F}$ & $7$ & $1$ & $1$ \\ 
${\bf A2}$ & $21$ & $5$ & $3$ \\
${\bf S2}$ & $28$ & $9$ & $11$ \\ \hline
${\bf Ra}$ & $140$ & $55$ & $1$ \\
${\bf A3}$ & $35$ & $10$ & $2$ \\
\end{tabular}
\hspace{0.5cm}
\begin{tabular}{l|c|c|c|c|}
	 $\SU(7)$ family  & $n_f$  & $77\ \xi_f$ & $n_g^f \leq$ & condition  \\\hline
${\bf 21}_{A2} +3\ {\bf \overline{7}}_F$ & $42$ & $8$ & $9\ (5)$ & \\ 
${\bf 28}_{S2} + 11\ {\bf \overline{7}}_F$ & $105$ & $20$ & $3\ (2)$ & \\
${\bf 28}_{S2}+ x\ {\bf \overline{21}}_{A2} + (11-3x)\ {\bf \overline{7}}_F$ & $105$ & $(20+2x)$ & $3\ (1)$ & $1\leq x \leq 2$\\
${\bf 28}_{S2}+ x\ {\bf \overline{21}}_{A2} + (11-3x)\ {\bf \overline{7}}_F$ & $105$ & $(20+2x)$ & $2\ (1)$ & $x =3$\\
${\bf 28}_{S2}+ 4\ {\bf \overline{21}}_{A2} + {\bf 7}_F$ & $119$ & $30$ & $2\ (1)$ & \\
$2\ {\bf 28}_{S2}+ 7\ {\bf \overline{21}}_{A2} + {\bf \overline{7}}_F$ & $210$ & $54$ & $1\ (0)$ & \\
\hline
${\bf 140}_{Ra} + {\bf \overline{7}}_F$ & $147$ & $56$ & $1\ (0)$ & \\
${\bf 140}_{Ra} + {\bf \overline{21}}_{A2} + 2\ {\bf 7}_{F}$ & $175$ & $62$ & $1\ (0)$ & \\
${\bf 140}_{Ra} + {\bf \overline{28}}_{S2} + x\ {\bf 21}_{A2} + (10-3x)\ {\bf 7}_{F}$ & $238$ & $(74+2x)$ & $1\ (0)$ & $0\leq x\leq 1$\\
\hline
${\bf 35}_{A3} + 2\ {\bf \overline{7}}_{F}$ & $49$ & $12$ & $6\ (3)$ & \\
${\bf 35}_{A3} + {\bf \overline{21}}_{A2} + {\bf 7}_{F}$ & $63$ & $16$ & $4\ (2)$ & \\
${\bf 35}_{A3} + {\bf \overline{28}}_{S2} + x\ {\bf 21}_{A2} + (9-3x)\ {\bf 7}_{F}$ & $126$ & $(30+2x)$ & $2\ (1)$ & $0\leq x\leq 3$\\\hline
$2\ {\bf 35}_{A3} + {\bf \overline{21}}_{A2} + {\bf \overline{7}}_{F}$ & $98$ & $26$ & $2\ (1)$ & \\
$2\ {\bf 35}_{A3} + {\bf \overline{28}}_{S2} + x\ {\bf 21}_{A2} + (7-3x)\ {\bf 7}_{F}$ & $147$ & $(36+2x)$ & $2\ (1)$ & $0\leq x\leq 1$\\
$2\ {\bf 35}_{A3} + {\bf \overline{28}}_{S2} + x\ {\bf 21}_{A2} + (7-3x)\ {\bf 7}_{F}$ & $147$ & $(36+2x)$ & $1\ (1)$ & $x=2 $\\\hline
$3\ {\bf 35}_{A3} + 2\ {\bf \overline{21}}_{A2}$ & $147$ & $40$ & $1\ (1)$ & \\
$3\ {\bf 35}_{A3} + {\bf \overline{28}}_{S2} + x\ {\bf \overline{21}}_{A2} + (5-3x)\ {\bf 7}_{F}$ & $168$ & $(44+2x)$ & $1\ (0)$ & $0\leq x\leq 1$\\
$3\ {\bf 35}_{A3} + {\bf \overline{28}}_{S2} + 2\ {\bf \overline{21}}_{A2} +  {\bf \overline{7}}_{F}$ & $182$ & $50$ & $1\ (0)$ & \\\hline
$4\ {\bf 35}_{A3} + {\bf \overline{28}}_{S2} + 3\ {\bf 7}_{F}$ & $189$ & $52$ & $1\ (0)$ & \\
$4\ {\bf 35}_{A3} + {\bf \overline{28}}_{S2} + \ {\bf 21}_{A2}$ & $189$ & $54$ & $1\ (0)$ & \\
$5\ {\bf 35}_{A3} + {\bf \overline{28}}_{S2} +  {\bf 7}_{F}$ & $210$ & $60$ & $1\ (0)$ & \\\hline
${\bf 140}_{Ra} + {\bf 35}_{A3} + {\bf \overline{21}}_{A2}$ & $196$ & $70$ & $1\ (0)$ & \\
${\bf 140}_{Ra} + {\bf \overline{35}}_{A3} + {\bf 7}_{F}$ & $182$ & $66$ & $1\ (0)$ & \\
\end{tabular}
	\caption{Complete list of  chiral families (right table) for $\SU(7)$, with representations labeled by their dimension.} \label{tab:su7}
\end{table}

\begin{table}[h]
	\centering
    	\begin{tabular}{l|c|c|c|}
	 Representation  & $d_r$  & $2T_r$ & $\mathcal{A}_r$   \\\hline
${\bf F}$ & $8$ & $1$ & $1$ \\ 
${\bf A2}$ & $28$ & $6$ & $4$ \\
${\bf S2}$ & $36$ & $10$ & $12$ \\ \hline
${\bf Ra}$ & $216$ & $75$ & $-3$ \\
${\bf A3}$ & $56$ & $15$ & $5$ \\
\end{tabular}
\hspace{0.5cm}
\begin{tabular}{l|c|c|c|c|}
	 $\SU(8)$ family  & $n_f$  & $88\ \xi_f$ & $n_g^f \leq$ & condition  \\\hline
${\bf 28}_{A2} +4\ {\bf \overline{8}}_F$ & $60$ & $10$ & $8\ (4)$ & \\ 
${\bf 36}_{S2} + 12\ {\bf \overline{8}}_F$ & $132$ & $22$ & $3\ (2)$ & \\
${\bf 36}_{S2}+ x\ {\bf \overline{28}}_{A2} + (12-4x)\ {\bf \overline{8}}_F$ & $132-4x$ & $(22+2x)$ & $3\ (1)$ & $1\leq x\leq 3$\\
\hline
${\bf 216}_{Ra} + 3\ {\bf 8}_F$ & $240$ & $78$ & $1\ (0)$ & \\
${\bf 216}_{Ra} + {\bf 28}_{A2} +  {\bf \overline{8}}_F$ & $252$ & $82$ & $1\ (0)$ & \\\hline
${\bf 56}_{A3} + 5\ {\bf \overline{8}}_{F}$ & $96$ & $20$ & $4\ (2)$ & \\
${\bf 56}_{A3} + {\bf \overline{28}}_{A2} +  {\bf \overline{8}}_{F}$ & $92$ & $22$ & $3\ (2)$ & \\
${\bf 56}_{A3} + 2\ {\bf \overline{28}}_{A2} +3\  {\bf 8}_{F}$ & $136$ & $30$ & $2\ (1)$ & \\
${\bf 56}_{A3} + {\bf \overline{36}}_{S2} + x\ {\bf 28}_{A2} + (7-4x)\  {\bf 8}_{F}$ & $148-4x$ & $(32+2x)$ & $2\ (1)$ & $0\leq x\leq 1$\\
${\bf 56}_{A3} + {\bf \overline{36}}_{S2} + 2\ {\bf 28}_{A2} +  {\bf \overline{8}}_{F}$ & $156$ & $38$ & $2\ (1)$ & \\\hline
$2\ {\bf 56}_{A3} + {\bf \overline{36}}_{S2} + 2\  {\bf 8}_{F}$ & $164$ & $42$ & $2\ (1)$ & \\
$2\ {\bf 56}_{A3} + {\bf \overline{36}}_{S2} + {\bf 28}_{A2} + 2\  {\bf \overline{8}}_{F}$ & $192$ & $48$ & $1\ (0)$ & \\\hline
$3\ {\bf 56}_{A3} + {\bf \overline{36}}_{S2} + 3\  {\bf \overline{8}}_{F}$ & $228$ & $58$ & $1\ (0)$ & \\
$3\ {\bf 56}_{A3} + {\bf \overline{36}}_{S2} + {\bf \overline{28}}_{A2} +  {\bf 8}_{F}$ & $240$ & $62$ & $1\ (0)$ & \\
\end{tabular}
	\caption{Complete list of  chiral families (right table) for $\SU(8)$, with representations labeled by their dimension.} \label{tab:su8}
\end{table}

\begin{table}[h]
	\centering
    	\begin{tabular}{l|c|c|c|}
	 Representation  & $d_r$  & $2T_r$ & $\mathcal{A}_r$   \\\hline
${\bf F}$ & $9$ & $1$ & $1$ \\ 
${\bf A2}$ & $36$ & $7$ & $5$ \\
${\bf S2}$ & $45$ & $11$ & $13$ \\ \hline
${\bf A4}$ & $126$ & $35$ & $5$ \\
${\bf A3}$ & $84$ & $21$ & $9$ \\
\end{tabular}
\hspace{0.5cm}
\begin{tabular}{l|c|c|c|c|}
	 $\SU(9)$ family  & $n_f$  & $99\ \xi_f$ & $n_g^f \leq$ & condition  \\\hline
${\bf 36}_{A2} +5\ {\bf \overline{9}}_F$ & $81$ & $12$ & $8\ (4)$ & \\ 
${\bf 45}_{S2} + 13\ {\bf \overline{9}}_F$ & $162$ & $24$ & $4\ (2)$ & \\
${\bf 45}_{S2} + {\bf \overline{36}}_{A2} + 8\ {\bf \overline{9}}_F$ & $153$ & $26$ & $3\ (2)$ &\\
${\bf 45}_{S2} +2\ {\bf \overline{36}}_{A2} + 3\ {\bf \overline{9}}_F$ & $144$ & $28$ & $3\ (1)$ & \\
${\bf 45}_{S2} +3\ {\bf \overline{36}}_{A2} + 2\ {\bf 9}_F$ & $171$ & $34$ & $2\ (1)$ & \\
$2\ {\bf 45}_{S2} +5\ {\bf \overline{36}}_{A2} + {\bf \overline{9}}_F$ & $279$ & $58$ & $1\ (0)$ & \\
$3\ {\bf 45}_{S2} +8\ {\bf \overline{36}}_{A2} + {\bf 9}_F$ & $432$ & $90$ & $1\ (0)$ & \\\hline
${\bf 126}_{A4} + 5\ {\bf \overline{9}}_F$ & $171$ & $40$ & $2\ (1)$ & \\
${\bf 126}_{A4} + {\bf \overline{36}}_{A2}$ & $162$ & $42$ & $2\ (1)$ & \\
${\bf 126}_{A4} + {\bf \overline{45}}_{S2} + x\ {\bf 36}_{A2} + (8-5x)\ {\bf 9}_F$ & $243-9x$ & $(54+2x)$ & $1\ (0)$ & $0\leq x\leq 1$\\
${\bf 126}_{A4} + {\bf \overline{45}}_{S2} + 2\ {\bf 36}_{A2} + 2\ {\bf \overline{9}}_F$ & $261$ & $62$ & $1\ (0)$ & \\ \hline
$2\ {\bf 126}_{A4} + {\bf \overline{45}}_{S2}  + 3\ {\bf 9}_F$ & $324$ & $84$ & $1\ (0)$ & \\
$2\ {\bf 126}_{A4} + {\bf \overline{45}}_{S2} + {\bf 36}_{A2} + 2\ {\bf \overline{9}}_F$ & $351$ & $90$ & $1\ (0)$ & \\\hline
${\bf 84}_{A3} + x\ {\bf \overline{36}}_{A2}  + (9-5x)\ {\bf \overline{9}}_F$ & $165-9x$ & $(30+2x)$ & $3\ (1)$ & $0\leq x\leq 1$\\
${\bf 84}_{A3} + 2\ {\bf \overline{36}}_{A2}  + {\bf 9}_F$ & $165$ & $36$ & $2\ (1)$ & \\
${\bf 84}_{A3} + {\bf \overline{45}}_{S2}  + 4\ {\bf 9}_F$ & $165$ & $36$ & $2\ (1)$ & \\
${\bf 84}_{A3} + {\bf \overline{45}}_{S2} + {\bf 36}_{A2}  + {\bf \overline{9}}_F$ & $174$ & $40$ & $2\ (1)$ & \\\hline
$2\ {\bf 84}_{A3} + {\bf \overline{45}}_{S2}  + 5\ {\bf \overline{9}}_F$ & $258$ & $58$ & $1\ (0)$ & \\
$2\ {\bf 84}_{A3} + {\bf \overline{45}}_{S2}  + {\bf \overline{36}}_{A2}$ & $249$ & $60$ & $1\ (0)$ & \\ \hline
${\bf 84}_{A3} + {\bf 126}_{A4} + {\bf \overline{45}}_{S2}  + {\bf \overline{9}}_F$ & $264$ & $68$ & $1\ (0)$ & \\
${\bf 84}_{A3} + {\bf \overline{126}}_{A4}  + 4\ {\bf \overline{9}}_F$ & $246$ & $60$ & $1\ (0)$ & \\
${\bf 84}_{A3} + {\bf \overline{126}}_{A4} +  {\bf \overline{36}}_{A2} + {\bf 9}_F$ & $255$ & $64$ & $1\ (0)$ & \\
${\bf 84}_{A3} + 2\ {\bf \overline{126}}_{A4}  + {\bf 9}_F$ & $345$ & $92$ & $1\ (0)$ & \\
\end{tabular}
	\caption{Complete list of  chiral families (right table) for $\SU(9)$, with representations labeled by their dimension.} \label{tab:su9}
\end{table}

\begin{table}[h]
	\centering
    	\begin{tabular}{l|c|c|c|}
	 Representation  & $d_r$  & $2T_r$ & $\mathcal{A}_r$   \\\hline
${\bf F}$ & $10$ & $1$ & $1$ \\ 
${\bf A2}$ & $45$ & $8$ & $6$ \\
${\bf S2}$ & $55$ & $12$ & $14$ \\ \hline
${\bf A3}$ & $120$ & $28$ & $14$ \\
${\bf A4}$ & $210$ & $56$ & $14$ \\
\end{tabular}
\hspace{0.5cm}
\begin{tabular}{l|c|c|c|c|}
	 $\SU(10)$ family  & $n_f$  & $110\ \xi_f$ & $n_g^f \leq$ & condition  \\\hline
${\bf 45}_{A2} +6\ {\bf \overline{10}}_F$ & $105$ & $14$ & $7\ (4)$ & \\ 
${\bf 55}_{S2}  + 14\ {\bf \overline{10}}_F$ & $195$ & $26$ & $4\ (2)$ & \\
${\bf 55}_{S2} + {\bf \overline{45}}_{A2} + 8\ {\bf \overline{10}}_F$ & $180$ & $28$ & $3\ (2)$ & \\
${\bf 55}_{S2} + 2\ {\bf \overline{45}}_{A2} + 2\ {\bf \overline{10}}_F$ & $165$ & $30$ & $3\ (1)$ & \\
${\bf 55}_{S2} + 3\ {\bf \overline{45}}_{A2} + 4\ {\bf 10}_F$ & $230$ & $40$ & $2\ (1)$ & \\
$2\ {\bf 55}_{S2} + 5\ {\bf \overline{45}}_{A2} + 2\ {\bf 10}_F$ & $355$ & $66$ & $1\ (0)$ & \\
$3\ {\bf 55}_{S2} + 7\ {\bf \overline{45}}_{A2}$ & $480$ & $92$ & $1\ (0)$ & \\
\hline
${\bf 120}_{A3} + x\ {\bf \overline{45}}_{A2} + (14-6x)\ {\bf \overline{10}}_F$ & $260-15x$ & $(42+2x)$ & $2\ (1)$ & $0\leq x\leq 2$ \\
${\bf 120}_{A3} + 3\ {\bf \overline{45}}_{A2} + 4\ {\bf 10}_F$ & $295$ & $56$ & $1\ (1)$ & \\
${\bf 120}_{A3} + {\bf \overline{55}}_{S2}$ & $175$ & $40$ & $2\ (1)$ &  \\
${\bf 120}_{A3} + {\bf 55}_{S2} + 5\ {\bf \overline{45}}_{A2} + 2\ {\bf 10}_{F}$ & $420$ & $82$ & $1\ (0)$ &  \\
${\bf 120}_{A3} + 2\ {\bf 55}_{S2} + 7\ {\bf \overline{45}}_{A2}$ & $545$ & $108$ & $1\ (0)$ &  \\
$2\ {\bf 120}_{A3} + 5\ {\bf \overline{45}}_{A2} + 2\ {\bf 10}_{F}$ & $485$ & $98$ & $1\ (0)$ &  \\\hline
${\bf 210}_{A4} + x\ {\bf \overline{45}}_{A2} + (14-6x)\ {\bf \overline{10}}_F$ & $350-15x$ & $(70+2x)$ & $1\ (0)$ & $0\leq x\leq 2$ \\
${\bf 210}_{A4} + 3\ {\bf \overline{45}}_{A2} + 4\ {\bf 10}_F$ & $385$ & $84$ & $1\ (0)$ & \\
${\bf 210}_{A4} + {\bf \overline{55}}_{S2}$ & $265$ & $68$ & $1\ (0)$ &  \\\hline
${\bf 210}_{A4} + {\bf \overline{120}}_{A3}$ & $330$ & $84$ & $1\ (0)$ & \\ 
\end{tabular}
	\caption{Complete list of  chiral families (right table) for $\SU(10)$, with representations labeled by their dimension.} \label{tab:su10}
\end{table}

\begin{table}[h]
	\centering
    	\begin{tabular}{l|c|c|c|}
	 Representation  & $d_r$  & $2T_r$ & $\mathcal{A}_r$   \\\hline
${\bf F}$ & $11$ & $1$ & $1$ \\ 
${\bf A2}$ & $55$ & $9$ & $7$ \\
${\bf S2}$ & $66$ & $13$ & $15$ \\ \hline
${\bf A3}$ & $165$ & $36$ & $20$ \\
${\bf A4}$ & $330$ & $84$ & $28$ \\
\end{tabular}
\hspace{0.5cm}
\begin{tabular}{l|c|c|c|c|}
	 $\SU(11)$ family  & $n_f$  & $121\ \xi_f$ & $n_g^f \leq$ & condition  \\\hline
${\bf 55}_{A2} +7\ {\bf \overline{11}}_F$ & $132$ & $16$ & $7\ (4)$ & \\ 
${\bf 66}_{S2} + x\ {\bf \overline{55}}_{A2}  + (15-7x)\ {\bf \overline{11}}_F$ & $231-22x$ & $(28+2x)$ & $4\ (2)$ & $0\leq x\leq 1$\\
${\bf 66}_{S2} + x\ {\bf \overline{55}}_{A2}  + (15-7x)\ {\bf \overline{11}}_F$ & $231-22x$ & $(28+2x)$ & $3\ (2)$ & $x=2$\\
${\bf 66}_{S2} + 3\ {\bf \overline{55}}_{A2}  + 6\ {\bf 11}_F$ & $297$ & $46$ & $2\ (1)$ &\\
$2\ {\bf 66}_{S2} + 3\ {\bf \overline{55}}_{A2}  + 9\ {\bf \overline{11}}_F$ & $396$ & $62$ & $1\ (1)$ &\\
$2\ {\bf 66}_{S2} + 5\ {\bf \overline{55}}_{A2}  + 5\ {\bf 11}_F$ & $462$ & $76$ & $1\ (0)$ &\\
$3\ {\bf 66}_{S2} + 7\ {\bf \overline{55}}_{A2}  + 4\ {\bf 11}_F$ & $627$ & $106$ & $1\ (0)$ &\\\hline
${\bf 165}_{A3} + x\ {\bf \overline{55}}_{A2} + (20-7x)\ {\bf \overline{11}}_F$ & $385-22x$ & $(56+2x)$ & $2\ (1)$ & $0\leq x\leq 2$ \\
${\bf 165}_{A3} + 3\ {\bf \overline{55}}_{A2} + {\bf 11}_F$ & $341$ & $64$ & $1\ (1)$ & \\
${\bf 165}_{A3} + {\bf \overline{66}}_{S2} + 5\ {\bf \overline{11}}_F$ & $286$ & $54$ & $2\ (1)$ & \\
${\bf 165}_{A3} + {\bf \overline{66}}_{S2} + {\bf \overline{55}}_{A2} + 2\ {\bf 11}_{F}$ & $308$ & $60$ & $2\ (1)$ & \\
${\bf 165}_{A3} + 2\ {\bf \overline{66}}_{S2} + x\ {\bf 55}_{A2} + (10-7x)\ {\bf 11}_{F}$ & $407-22x$ & $(72+2x)$ & $1\ (0)$ & $0\leq x\leq 1$\\
${\bf 165}_{A3} + 2\ {\bf \overline{66}}_{S2} + 2\ {\bf 55}_{A2} + 4\ {\bf \overline{11}}_{F}$ & $451$ & $84$ & $1\ (0)$ & \\
${\bf 165}_{A3} + 3\ {\bf \overline{66}}_{S2} + 4\ {\bf 55}_{A2} + 3\ {\bf \overline{11}}_{F}$ & $616$ & $114$ & $1\ (0)$ & \\\hline
${\bf 330}_{A4} + x\ {\bf \overline{55}}_{A2} + (28-7x)\ {\bf \overline{11}}_{F}$ & $638-22x$ & $(112+2x)$ & $1\ (0)$ & $0\leq x\leq 4$\\
${\bf 330}_{A4} + {\bf \overline{66}}_{S2}+ x\ {\bf \overline{55}}_{A2} + (13-7x)\ {\bf \overline{11}}_{F}$ & $539-22x$ & $(110+2x)$ & $1\ (0)$ & $0\leq x\leq 1$\\
${\bf 330}_{A4} + {\bf \overline{66}}_{S2}+ 2\ {\bf \overline{55}}_{A2} +  {\bf 11}_{F}$ & $517$ & $116$ & $1\ (0)$ & \\
${\bf 330}_{A4} + 2\ {\bf \overline{66}}_{S2}+ 2\  {\bf 11}_{F}$ & $484$ & $112$ & $1\ (0)$ & \\
\end{tabular}
	\caption{Complete list of  chiral families (right table) for $\SU(11)$, with representations labeled by their dimension.} \label{tab:su11}
\end{table}

\begin{table}[h]
	\centering
    	\begin{tabular}{l|c|c|c|}
	 Representation  & $d_r$  & $2T_r$ & $\mathcal{A}_r$   \\\hline
${\bf F}$ & $12$ & $1$ & $1$ \\ 
${\bf A2}$ & $66$ & $10$ & $8$ \\
${\bf S2}$ & $78$ & $14$ & $16$ \\ \hline
${\bf A3}$ & $220$ & $45$ & $27$ \\
\end{tabular}
\hspace{0.5cm}
\begin{tabular}{l|c|c|c|c|}
	 $\SU(12)$ family  & $n_f$  & $132\ \xi_f$ & $n_g^f \leq$ & condition  \\\hline
${\bf 66}_{A2} +8\ {\bf \overline{12}}_F$ & $162$ & $18$ & $7\ (3)$ & \\ 
${\bf 78}_{S2} + x\ {\bf \overline{66}}_{A2}  + (16-8x)\ {\bf \overline{12}}_F$ & $270-30x$ & $(30+2x)$ & $4\ (2)$ & $0\leq x\leq 1$\\
${\bf 78}_{S2} + 2\ {\bf \overline{66}}_{A2} $ & $210$ & $34$ & $3\ (2)$ & \\\hline
${\bf 220}_{A3} + x\ {\bf \overline{66}}_{A2} + (27-8x)\ {\bf \overline{12}}_F$ & $544-30x$ & $(72+2x)$ & $1\ (0)$ & $0\leq x\leq 3$ \\
${\bf 220}_{A3} + 4\ {\bf \overline{66}}_{A2} + 5\ {\bf 12}_F$ & $544$ & $90$ & $1\ (0)$ &  \\
${\bf 220}_{A3} + {\bf \overline{78}}_{S2}  + 11\ {\bf \overline{12}}_F$ & $430$ & $70$ & $1\ (1)$ &  \\
${\bf 220}_{A3} + {\bf \overline{78}}_{S2} +  {\bf \overline{66}}_{A2} + 3\ {\bf \overline{12}}_F$ & $400$ & $72$ & $1\ (0)$ & \\
${\bf 220}_{A3} +  {\bf \overline{78}}_{S2} + 2\ {\bf \overline{66}}_{A2} + 5\ {\bf 12}_F$ & $490$ & $84$ & $1\ (0)$ &  \\
${\bf 220}_{A3} + 2\ {\bf \overline{78}}_{S2} + 5\ {\bf 12}_F$ & $436$ & $78$ & $1\ (0)$ &  \\
${\bf 220}_{A3} + 2\ {\bf \overline{78}}_{S2} + {\bf 66}_{A2} + 3\ {\bf \overline{12}}_{F}$ & $478$ & $86$ & $1\ (0)$ &  \\
\end{tabular}
	\caption{Complete list of  chiral families (right table) for $\SU(12)$, with representations labeled by their dimension.} \label{tab:su12}
\end{table}

\begin{table}[h]
	\centering
    	\begin{tabular}{l|c|c|c|}
	 Representation  & $d_r$  & $2T_r$ & $\mathcal{A}_r$   \\\hline
${\bf F}$ & $13$ & $1$ & $1$ \\ 
${\bf A2}$ & $78$ & $11$ & $9$ \\
${\bf S2}$ & $91$ & $15$ & $17$ \\ \hline
${\bf A3}$ & $286$ & $55$ & $35$ \\
\end{tabular}
\hspace{0.5cm}
\begin{tabular}{l|c|c|c|c|}
	 $\SU(13)$ family  & $n_f$  & $143\ \xi_f$ & $n_g^f \leq$ & condition  \\\hline
${\bf 78}_{A2} +9\ {\bf \overline{13}}_F$ & $195$ & $20$ & $7\ (3)$ & \\ 
${\bf 91}_{S2} + x\ {\bf \overline{78}}_{A2}  + (17-9x)\ {\bf \overline{13}}_F$ & $312-39x$ & $(32+2x)$ & $4\ (2)$ & $0\leq x\leq 1$\\
${\bf 91}_{S2} + 2\ {\bf \overline{78}}_{A2}  +{\bf 13}_F$ & $260$ & $38$ & $3\ (2)$ & \\
$2\ {\bf 91}_{S2} + 3\ {\bf \overline{78}}_{A2}  +7\ {\bf \overline{13}}_F$ & $507$ & $70$ & $2\ (1)$ & \\
$3\ {\bf 91}_{S2} + 5\ {\bf \overline{78}}_{A2}  +6\ {\bf \overline{13}}_F$ & $741$ & $106$ & $1\ (0)$ & \\
$4\ {\bf 91}_{S2} + 7\ {\bf \overline{78}}_{A2}  +5\ {\bf \overline{13}}_F$ & $975$ & $142$ & $1\ (0)$ & \\\hline
${\bf 286}_{A3} + x\ {\bf \overline{78}}_{A2} + (35-9x)\ {\bf \overline{13}}_F$ & $741-39x$ & $(90+2x)$ & $1\ (0)$ & $0\leq x\leq 3$ \\
${\bf 286}_{A3} + 4\ {\bf \overline{78}}_{A2} + {\bf 13}_F$ & $611$ & $100$ & $1\ (0)$ &  \\
${\bf 286}_{A3} + {\bf \overline{91}}_{S2}+ x\ {\bf \overline{78}}_{A2} + (18-9x)\ {\bf \overline{13}}_F$ & $611-39x$ & $(88+2x)$ & $1\ (0)$ & $0\leq x\leq 2$ \\
${\bf 286}_{A3} + 2\ {\bf \overline{91}}_{S2}+  {\bf \overline{13}}_F$ & $481$ & $86$ & $1\ (0)$ & \\
${\bf 286}_{A3} + 2\ {\bf \overline{91}}_{S2}+  {\bf \overline{78}}_{A2} + 8\ {\bf 13}_F$ & $650$ & $104$ & $1\ (0)$ & \\
${\bf 286}_{A3} + 3\ {\bf \overline{91}}_{S2} + x\ {\bf 78}_{A2} + (16-9x)\ {\bf 13}_F$ & $767-39x$ & $(116+2x)$ & $1\ (0)$ & $0\leq x\leq 1$\\
${\bf 286}_{A3} + 3\ {\bf \overline{91}}_{S2} + 2\ {\bf 78}_{A2} + 2\ {\bf \overline{13}}_F$ & $741$ & $124$ & $1\ (0)$ & \\
\end{tabular}
	\caption{Complete list of  chiral families (right table) for $\SU(13)$, with representations labeled by their dimension.} \label{tab:su13}
\end{table}

\begin{table}[h]
	\centering
    	\begin{tabular}{l|c|c|c|}
	 Representation  & $d_r$  & $2T_r$ & $\mathcal{A}_r$   \\\hline
${\bf F}$ & $14$ & $1$ & $1$ \\ 
${\bf A2}$ & $91$ & $12$ & $10$ \\
${\bf S2}$ & $105$ & $16$ & $18$ \\ \hline
${\bf A3}$ & $364$ & $66$ & $44$ \\
\end{tabular}
\hspace{0.5cm}
\begin{tabular}{l|c|c|c|c|}
	 $\SU(14)$ family  & $n_f$  & $154\ \xi_f$ & $n_g^f \leq$ & condition  \\\hline
${\bf 91}_{A2} +10\ {\bf \overline{14}}_F$ & $231$ & $22$ & $6\ (3)$ & \\ 
${\bf 105}_{S2} + x\ {\bf \overline{91}}_{A2}  + (18-10x)\ {\bf \overline{14}}_F$ & $357-49x$ & $(34+2x)$ & $4\ (2)$ & $0\leq x\leq 1$\\
${\bf 105}_{S2} + 2\ {\bf \overline{91}}_{A2}  +2\ {\bf 14}_F$ & $315$ & $42$ & $3\ (1)$ & \\
$2\ {\bf 105}_{S2} + 3\ {\bf \overline{91}}_{A2}  +6\ {\bf \overline{14}}_F$ & $567$ & $74$ & $2\ (1)$ & \\
$3\ {\bf 105}_{S2} + 5\ {\bf \overline{91}}_{A2}  +4\ {\bf \overline{14}}_F$ & $826$ & $112$ & $1\ (0)$ & \\
$4\ {\bf 105}_{S2} + 7\ {\bf \overline{91}}_{A2}  +2\ {\bf \overline{14}}_F$ & $1085$ & $150$ & $1\ (0)$ & \\\hline
${\bf 364}_{A3} + x\ {\bf \overline{91}}_{A2} + (44-10x)\ {\bf \overline{14}}_F$ & $980-49x$ & $(110+2x)$ & $1\ (0)$ & $0\leq x\leq 4$ \\
${\bf 364}_{A3} + 5\ {\bf \overline{91}}_{A2} + 6\ {\bf 14}_F$ & $903$ & $132$ & $1\ (0)$ &  \\
${\bf 364}_{A3} + {\bf \overline{105}}_{S2}+ x\ {\bf \overline{91}}_{A2} + (26-10x)\ {\bf \overline{14}}_F$ & $833-49x$ & $(108+2x)$ & $1\ (0)$ & $0\leq x\leq 2$ \\
${\bf 364}_{A3} + {\bf \overline{105}}_{S2}+ 3\ {\bf \overline{91}}_{A2} + 4\ {\bf 14}_F$ & $798$ & $122$ & $1\ (0)$ & \\
${\bf 364}_{A3} + 2\ {\bf \overline{105}}_{S2}+ 8\ {\bf \overline{14}}_F$ & $686$ & $106$ & $1\ (0)$ & \\
${\bf 364}_{A3} + 2\ {\bf \overline{105}}_{S2}+  {\bf \overline{91}}_{A2} + 2\ {\bf 14}_F$ & $693$ & $112$ & $1\ (0)$ & \\
${\bf 364}_{A3} + 3\ {\bf \overline{105}}_{S2} + 10\ {\bf 14}_F$ & $819$ & $124$ & $1\ (0)$ & \\
${\bf 364}_{A3} + 3\ {\bf \overline{105}}_{S2} +  {\bf 91}_{A2} $ & $770$ & $126$ & $1\ (0)$ & \\
\end{tabular}
	\caption{Complete list of  chiral families (right table) for $\SU(14)$, with representations labeled by their dimension.} \label{tab:su14}
\end{table}

\begin{table}[h]
	\centering
    	\begin{tabular}{l|c|c|c|}
	 Representation  & $d_r$  & $2T_r$ & $\mathcal{A}_r$   \\\hline
${\bf F}$ & $15$ & $1$ & $1$ \\ 
${\bf A2}$ & $105$ & $13$ & $11$ \\
${\bf S2}$ & $120$ & $17$ & $19$ \\ \hline
${\bf A3}$ & $455$ & $78$ & $54$ \\
\end{tabular}
\hspace{0.5cm}
\begin{tabular}{l|c|c|c|c|}
	 $\SU(15)$ family  & $n_f$  & $165\ \xi_f$ & $n_g^f \leq$ & condition  \\\hline
${\bf 105}_{A2} +11\ {\bf \overline{15}}_F$ & $270$ & $24$ & $6\ (3)$ & \\ 
${\bf 120}_{S2} + x\ {\bf \overline{105}}_{A2}  + (19-11x)\ {\bf \overline{15}}_F$ & $405-60x$ & $(36+2x)$ & $4\ (2)$ & $0\leq x\leq 1$\\
${\bf 120}_{S2} + 2\ {\bf \overline{105}}_{A2}  + 3\ {\bf 15}_F$ & $375$ & $46$ & $3\ (1)$ & \\
$2\ {\bf 120}_{S2} + 3\ {\bf \overline{105}}_{A2}  +5\ {\bf \overline{15}}_F$ & $630$ & $78$ & $2\ (1)$ & \\
$3\ {\bf 120}_{S2} + 5\ {\bf \overline{105}}_{A2}  +2\ {\bf \overline{12}}_F$ & $915$ & $118$ & $1\ (0)$ & \\
$4\ {\bf 120}_{S2} + 7\ {\bf \overline{105}}_{A2}  + {\bf 15}_F$ & $1230$ & $160$ & $1\ (0)$ & \\\hline
${\bf 455}_{A3} + x\ {\bf \overline{105}}_{A2} + (54-11x)\ {\bf \overline{15}}_F$ & $1265-60x$ & $(132+2x)$ & $1\ (0)$ & $0\leq x\leq 4$ \\
${\bf 455}_{A3} + 5\ {\bf \overline{105}}_{A2} + {\bf 15}_F$ & $995$ & $144$ & $1\ (0)$ &  \\
${\bf 455}_{A3} + {\bf \overline{120}}_{S2}+ x\ {\bf \overline{105}}_{A2} + (35-11x)\ {\bf \overline{15}}_F$ & $1100-60x$ & $(130+2x)$ & $1\ (0)$ & $0\leq x\leq 3$ \\
${\bf 455}_{A3} + {\bf \overline{120}}_{S2}+ 4\ {\bf \overline{105}}_{A2} + 9\ {\bf 15}_F$ & $1130$ & $156$ & $1\ (0)$ & \\
${\bf 455}_{A3} + 2\ {\bf \overline{120}}_{S2}+ x\ {\bf \overline{105}}_{A2} + (16-11x)\ {\bf \overline{15}}_F$ & $935-60x$ & $(128+2x)$ & $1\ (0)$ & $0\leq x\leq 1$\\
${\bf 455}_{A3} + 2\ {\bf \overline{120}}_{S2}+  2\ {\bf \overline{105}}_{A2} + 6\ {\bf 15}_F$ & $995$ & $144$ & $1\ (0)$ & \\
${\bf 455}_{A3} + 3\ {\bf \overline{120}}_{S2} + 3\ {\bf 15}_F$ & $860$ & $132$ & $1\ (0)$ & \\
${\bf 455}_{A3} + 3\ {\bf \overline{120}}_{S2} + {\bf 105}_{A2}+ 8\ {\bf \overline{15}}_F$ & $1040$ & $150$ & $1\ (0)$ & \\
\end{tabular}
	\caption{Complete list of  chiral families (right table) for $\SU(15)$, with representations labeled by their dimension.} \label{tab:su15}
\end{table}

\begin{table}[h]
	\centering
    	\begin{tabular}{l|c|c|c|}
	 Representation  & $d_r$  & $2T_r$ & $\mathcal{A}_r$   \\\hline
${\bf F}$ & $16$ & $1$ & $1$ \\ 
${\bf A2}$ & $120$ & $14$ & $12$ \\
${\bf S2}$ & $136$ & $18$ & $20$ \\ \hline
${\bf A3}$ & $560$ & $91$ & $65$ \\
\end{tabular}
\hspace{0.5cm}
\begin{tabular}{l|c|c|c|c|}
	 $\SU(16)$ family  & $n_f$  & $176\ \xi_f$ & $n_g^f \leq$ & condition  \\\hline
${\bf 120}_{A2} +12\ {\bf \overline{16}}_F$ & $312$ & $26$ & $6\ (3)$ & \\ 
${\bf 136}_{S2} + x\ {\bf \overline{120}}_{A2}  + (20-12x)\ {\bf \overline{16}}_F$ & $456-72x$ & $(38+2x)$ & $4\ (2)$ & $0\leq x\leq 1$\\
${\bf 136}_{S2} + 2\ {\bf \overline{120}}_{A2}  + 4\ {\bf 16}_F$ & $440$ & $50$ & $3\ (1)$ & \\
$2\ {\bf 136}_{S2} + 3\ {\bf \overline{120}}_{A2}  +4\ {\bf \overline{16}}_F$ & $696$ & $82$ & $2\ (1)$ & \\
$3\ {\bf 136}_{S2} + 5\ {\bf \overline{120}}_{A2} $ & $1008$ & $124$ & $1\ (0)$ & \\
\hline
${\bf 560}_{A3} + x\ {\bf \overline{120}}_{A2} + (65-12x)\ {\bf \overline{16}}_F$ & $1600-72x$ & $(156+2x)$ & $1\ (0)$ & $0\leq x\leq 5$ \\
${\bf 560}_{A3} + {\bf \overline{136}}_{S2}+ x\ {\bf \overline{120}}_{A2} + (45-12x)\ {\bf \overline{16}}_F$ & $1416-72x$ & $(154+2x)$ & $1\ (0)$ & $0\leq x\leq 3$ \\
${\bf 560}_{A3} + {\bf \overline{136}}_{S2}+ 4\ {\bf \overline{120}}_{A2} + 3\ {\bf 16}_F$ & $1224$ & $168$ & $1\ (0)$ & \\
${\bf 560}_{A3} + 2\ {\bf \overline{136}}_{S2}+ x\ {\bf \overline{120}}_{A2} + (25-12x)\ {\bf \overline{16}}_F$ & $1232-72x$ & $(152+2x)$ & $1\ (0)$ & $0\leq x\leq 2$\\

${\bf 560}_{A3} + 3\ {\bf \overline{136}}_{S2} + 5\ {\bf \overline{16}}_F$ & $1048$ & $150$ & $1\ (0)$ & \\
${\bf 560}_{A3} + 3\ {\bf \overline{136}}_{S2} + {\bf \overline{120}}_{A2}+ 7\ {\bf 16}_F$ & $1200$ & $166$ & $1\ (0)$ & \\
\end{tabular}
	\caption{Complete list of  chiral families (right table) for $\SU(16)$, with representations labeled by their dimension.} \label{tab:su16}
\end{table}

\begin{table}[h]
	\centering
    	\begin{tabular}{l|c|c|c|}
	 Representation  & $d_r$  & $2T_r$ & $\mathcal{A}_r$   \\\hline
${\bf F}$ & $17$ & $1$ & $1$ \\ 
${\bf A2}$ & $136$ & $15$ & $13$ \\
${\bf S2}$ & $153$ & $19$ & $21$ \\ \hline
${\bf A3}$ & $680$ & $105$ & $77$ \\
\end{tabular}
\hspace{0.5cm}
\begin{tabular}{l|c|c|c|c|}
	 $\SU(17)$ family  & $n_f$  & $187\ \xi_f$ & $n_g^f \leq$ & condition  \\\hline
${\bf 136}_{A2} +13\ {\bf \overline{17}}_F$ & $357$ & $28$ & $6\ (3)$ & \\ 
${\bf 153}_{S2} + x\ {\bf \overline{136}}_{A2}  + (21-13x)\ {\bf \overline{17}}_F$ & $510-85x$ & $(40+2x)$ & $4\ (2)$ & $0\leq x\leq 1$\\
${\bf 153}_{S2} + 2\ {\bf \overline{136}}_{A2}  + 5\ {\bf 17}_F$ & $510$ & $54$ & $3\ (1)$ & \\
$2\ {\bf 153}_{S2} + 3\ {\bf \overline{136}}_{A2}  +3\ {\bf \overline{17}}_F$ & $765$ & $86$ & $2\ (1)$ & \\
$3\ {\bf 153}_{S2} + 5\ {\bf \overline{136}}_{A2} + 2\ {\bf 17}_{F}$ & $1173$ & $134$ & $1\ (0)$ & \\
\hline
${\bf 680}_{A3} + x\ {\bf \overline{136}}_{A2} + (77-13x)\ {\bf \overline{17}}_F$ & $1989-85x$ & $(182+2x)$ & $1\ (0)$ & $0\leq x\leq 2$ \\
${\bf 680}_{A3} + {\bf \overline{153}}_{S2}+ x\ {\bf \overline{136}}_{A2} + (56-13x)\ {\bf \overline{17}}_F$ & $1785-85x$ & $(180+2x)$ & $1\ (0)$ & $0\leq x\leq 3$ \\
${\bf 680}_{A3} + 2\ {\bf \overline{153}}_{S2}+ x\ {\bf \overline{136}}_{A2} + (35-13x)\ {\bf \overline{17}}_F$ & $1581-85x$ & $(178+2x)$ & $1\ (0)$ & $0\leq x\leq 2$ \\
${\bf 680}_{A3} + 3\ {\bf \overline{153}}_{S2}+ x\ {\bf \overline{136}}_{A2} + (14-13x)\ {\bf \overline{17}}_F$ & $1377-85x$ & $(176+2x)$ & $1\ (0)$ & $0\leq x\leq 1$ \\
\end{tabular}
	\caption{Complete list of  chiral families (right table) for $\SU(17)$, with representations labeled by their dimension.} \label{tab:su17}
\end{table}

\end{document}